\documentclass[prd, aps, nofootinbib, preprintnumbers, showpacs,superscriptaddress]{revtex4}

\usepackage{graphicx,epsfig}

\usepackage{amssymb,amsmath}
\usepackage[english]{babel}
\usepackage{graphicx}
\usepackage[latin9]{inputenc}
\usepackage{epstopdf}
\usepackage{color}

\def\ii{{\rm i}}

\newcommand{\be}{\begin{equation}}
\newcommand{\ee}{\end{equation}}

\begin{document}

\author{Ernst Nils Dorband}
\email{dorband@cct.lsu.edu}

\affiliation{Department of Physics and Astronomy, 202 Nicholson Hall,
  Louisiana State University, Baton Rouge, LA 70803, USA}
\homepage{http://relativity.phys.lsu.edu/}

\affiliation{Center for Computation and Technology, 302 Johnston Hall,
  Louisiana State University, Baton Rouge, LA 70803, USA}
\homepage{http://www.cct.lsu.edu/}

\author{Emanuele Berti}
\email{berti@wugrav.wustl.edu}

\affiliation{McDonnell Center for the Space Sciences, Department of
Physics, Washington University, St.\ Louis, MO 63130, USA}

\author{Peter Diener}
\email{diener@cct.lsu.edu}

\affiliation{Department of Physics and Astronomy, 202 Nicholson Hall,
  Louisiana State University, Baton Rouge, LA 70803, USA}
\homepage{http://relativity.phys.lsu.edu/}

\affiliation{Center for Computation and Technology, 302 Johnston Hall,
  Louisiana State University, Baton Rouge, LA 70803, USA}
\homepage{http://www.cct.lsu.edu/}

\author{Erik Schnetter}
\email{schnetter@cct.lsu.edu}

\affiliation{Center for Computation and Technology, 302 Johnston Hall,
  Louisiana State University, Baton Rouge, LA 70803, USA}
\homepage{http://www.cct.lsu.edu/}

\author{Manuel Tiglio}
\email{tiglio@cct.lsu.edu}

\affiliation{Department of Physics and Astronomy, 202 Nicholson Hall,
  Louisiana State University, Baton Rouge, LA 70803, USA}
\homepage{http://relativity.phys.lsu.edu/}

\affiliation{Center for Computation and Technology, 302 Johnston Hall,
  Louisiana State University, Baton Rouge, LA 70803, USA}
\homepage{http://www.cct.lsu.edu/}

\title{A numerical study of the quasinormal mode excitation of Kerr
  black holes}

 
 \preprint{LSU-REL-081806}
 
\begin{abstract}
  We present numerical results from three-dimensional evolutions of scalar
  perturbations of Kerr black holes. Our simulations make use of a high-order
  accurate multi-block code which naturally allows for fixed adaptivity and
  smooth inner (excision) and outer boundaries.  We focus on the quasinormal
  ringing phase, presenting a systematic method for extraction of the
  quasinormal mode frequencies and amplitudes and comparing our results
  against perturbation theory. 

  The detection of a single mode in a ringdown waveform allows for a
  measurement of the mass and spin of a black hole; a multimode detection
  would allow a test of the Kerr nature of the source.  Since the possibility
  of a multimode detection depends on the relative mode amplitude, we study
  this topic in some detail.
  The amplitude of each mode depends exponentially on the starting time of the
  quasinormal regime, which is not defined unambiguously. We show that this
  {\it time-shift problem} can be circumvented by looking at appropriately
  chosen {\em relative} mode amplitudes. From our simulations we extract the
  quasinormal frequencies and the relative and absolute amplitudes of
  corotating and counterrotating modes (including overtones in the corotating
  case). We study the dependence of these amplitudes on the shape of the
  initial perturbation, the angular dependence of the mode and the black hole
  spin, comparing against results from perturbation theory in the so-called
  asymptotic approximation. We also compare the quasinormal frequencies from
  our numerical simulations with predictions from perturbation theory, finding
  excellent agreement. For rapidly rotating black holes (of spin $j=0.98$) we
  can extract the quasinormal frequencies of not only the fundamental mode,
  but also of the first two overtones. Finally we study under what conditions
  the relative amplitude between given pairs of modes gets maximally excited
  and present a quantitative analysis of rotational mode--mode coupling. The
  main conclusions and techniques of our analysis are quite general and, as
  such, should be of interest in the study of ringdown gravitational waves
  produced by astrophysical gravitational wave sources.
\end{abstract}

\maketitle

%\tableofcontents

%%%%%%%%%%%%%%%%%%%%%%%%%%%%%%%%%%%%%%%%%%%%%
\section{Introduction}
%%%%%%%%%%%%%%%%%%%%%%%%%%%%%%%%%%%%%%%%%%%%%

One of the most useful methods to explore the response of black holes
to external perturbations is based on wave scattering
\cite{Futterman88}. Early studies identified three main stages in the
dynamics of a wave propagating on a black hole background, as observed
at a fixed spatial point. In a first, transient phase the observed
wave depends on the structure of the initial pulse. Vishveshwara and
Press discovered that this initial ``burst'' is invariably followed by
a second phase characterized by exponentially decaying oscillations: 
this phase is usually referred to as
``quasinormal ringing'' \cite{Vishveshwara70b,Press71}.  In the third
and last stage of the evolution, waves slowly die off as a power law
tail \cite{Price72}.

Astrophysical black holes should be well described by the Kerr solution,
since charge is unlikely to play a major role in astrophysical
scenarios (see e.g.\ \cite{Berti:2005eb} for a discussion).  As a
consequence of the ``no hair theorem'', if general
relativity is the correct theory of gravity, the quasinormal mode (QNM)
frequencies of a Kerr black hole depend only on its mass and angular
momentum. Earth-based and space-based gravitational wave detectors
have the potential to measure the frequency and damping time of a
QNM. From these two observables we can infer the black hole's mass and
angular momentum \cite{Echeverria89,Finn92}.  For the space-based {\it
Laser Interferometer Space Antenna} (LISA) , and possibly also for
second-generation ground-based detectors, the signal-to-noise ratio
can be large enough that we will be able to identify {\it two} or more
QNM frequencies in the signal \cite{Berti06b}.  A multi-mode detection
would provide a striking, direct test of the Kerr nature of the source
(i.e., of the no-hair theorem). The basic idea is quite
simple. Roughly speaking, the first mode in the pair is used to
determine the black hole's mass and angular momentum, and the other
mode(s) to verify that the QNM spectrum is indeed consistent with a
general relativistic Kerr black hole \cite{Dreyer04}.

The combined observation of supermassive black hole binary inspiral
and ringdown with LISA can provide even more information
\cite{Flanagan:1998a}.  Parameter estimation during the inspiral phase
can be very accurate, depending on the black holes' masses, spins and
distance \cite{Berti05a}. Combining information from the inspiral and
ringdown phases we can estimate the energy radiated in the merger, and
possibly improve parameter estimation from both phases (see
e.g.\ \cite{Luna06} for a preliminary study of this effect in the
context of earth-based detectors). 

In the last thirty years the development of gravitational wave astronomy
motivated a detailed investigation of the QNM frequency spectrum
\cite{Kokkotas99b,Nollert99,Berti04a}. In comparison, the problem of the {\it
  relative excitation} of QNMs received very little attention (see e.g.
\cite{Berti:2006wq} and references therein).  Ideally, the relative QNM
excitation should be determined by general relativistic simulations of binary
black hole mergers. Despite recent progress, this information is not yet
available \cite{Berti06a}.  Given the recent progress of numerical relativity,
by the time LISA flies we could have a good knowledge of the multipolar
distribution of the energy and angular momentum radiated in a black hole
merger under generic conditions. Knowing in advance which modes should be
excited in a realistic merger will not only be useful to probe the Kerr nature
of the source, but also to reduce the number of templates needed to perform
matched filtering on ringdown waveforms.

In this paper we present a quantitative investigation of QNM excitation
studying a simple model problem: the scattering of scalar waves on a Kerr
background. We use our new infrastructure for multi-block simulations
\cite{Schnetter06a} for these studies. The infrastructure is based on the
techniques described and applied in the context of numerical relativity in
\cite{Lehner2005a} and further extended in \cite{Diener05b}.  Our
infrastructure uses Carpet \cite{Schnetter-etal-03b, carpetweb}, a driver for
the Cactus computational toolkit \cite{Goodale02a, cactusweb1}, originally
designed to provide fixed and adaptive mesh refinement. The capabilities of
Carpet have been recently extended \cite{Schnetter06a} to include the type of
boundary conditions that are needed for multi-block (also called multi-patch)
simulations in Cactus.

Multi-block techniques yield increased efficiency and accuracy in our studies,
for two main reasons. The first is that we can set up smooth excision and
outer boundaries, and we can therefore apply boundary conditions in a clean
and well understood way.  Furthermore, a suitable multi-block grid structure
provides a natural and flexible way of implementing mesh refinement. We can
keep a fixed angular resolution throughout the entire domain, avoiding the
unnecessary high resolution at large distances from the central object that
one would have using a cartesian grid. The resources thus saved can be used,
for example, to set up a rather large number of grid points in the radial
direction, so that outer boundaries are located at large radii and the noise
produced at the boundaries does not affect the results. Even though we are working in
three dimensions, our results are more accurate than previous studies using
two-dimensional codes \cite{Krivan96a,Krivan97a}. Krivan {\it et
  al.}~\cite{Krivan96a} studied the late time dynamics and the rotational
coupling of massless scalar fields in a Kerr background, but not their
quasinormal ringing.  Later they extended the analysis to gravitational
perturbations, considering both the late time tail and the quasinormal ringing
phase~\cite{Krivan97a}.  For large rotation the damping times of corotating
fundamental modes in \cite{Krivan97a} are accurate within $\sim 3\%$ when
compared to results from perturbation theory; our accuracy ($\sim
0.3 \%$) is roughly an order of magnitude better. In fact, we can extract the
frequencies of some {\em overtones} with an error of the order of a few
percent or less.

Given the high accuracy of our multi-block infrastructure, a careful
extraction of the QNM content of the waveforms becomes necessary. We discuss
in detail the so-called {\it time-shift problem} (exponential dependence of the
quasinormal amplitudes on the time at which the quasinormal ringing regime
starts), how it affects the determination of both absolute and relative QNM
amplitudes, and how to choose pairs of modes so as to decrease the uncertainty
on relative amplitudes. We also introduce a general criterion (based on
minimizing a suitably defined residual) to determine the optimal fitting
window to extract QNM frequencies and amplitudes. Using these tools we study
the absolute and relative amplitudes of corotating and counterrotating modes
for Gaussian initial data located in the far zone. We study the dependence of
these amplitudes on the radial shape of the initial data, finding excellent
agreement with results from perturbation theory \cite{Berti:2006wq}.  We also
discuss the problem of extracting overtones for modes with a given angular
dependence, finding that the first overtones of corotating modes (e.g.\ modes
with $l=m=2$) contribute significantly to the waveform for rapidly rotating
black holes.

The plan of the paper is as follows. In Sec.~\ref{we} we briefly describe our
multi-block code. After introducing the background metric, we discuss the
numerical implementation of the scalar wave equation and our time evolution
techniques. In Sec.~\ref{QNM} we extract QNM frequencies and amplitudes from
our evolutions, comparing with analytical predictions from perturbation
theory. To start with, we point out some conceptual limitations in the
extraction of QNM amplitudes due to the so-called {\it time-shift problem}. Then we
introduce a rather general method to determine the best fitting interval to
extract QNM waveforms. We first check the accuracy of this method (and of our
numerical code) by reproducing the QNM frequencies predicted by standard
perturbation theory. Scalar QNM frequencies for Kerr black holes have been
computed in \cite{Berti:2005eb}, and they have never been systematically
confirmed by numerical time evolutions\footnote{See however \cite{Saijo97a},
  where the fundamental scalar mode with $l=0$ was observed to dominate the
  emission of scalar radiation by perturbed Kerr black holes in the
  Brans-Dicke theory of gravity.}.  Next we give a quantitative estimate of
rotational mode mixing as a function of the black hole's spin and discuss the
initial data dependence of the amplitudes of corotating and counterrotating
modes. In this way we assess the validity of the amplitudes predicted by
perturbation theory in the so-called asymptotic approximation (where both the
observer and the initial data are located far away from the black hole).
Finally we discuss the extraction of overtones from our waveforms.

\section{Numerical evolution of scalar perturbations of Kerr black holes}
\label{we}

\subsection{Grid structure}

\begin{figure*}[htb]
% \begin{center}
%   \begin{tabular}{cc}
\includegraphics[width=17cm,angle=0]{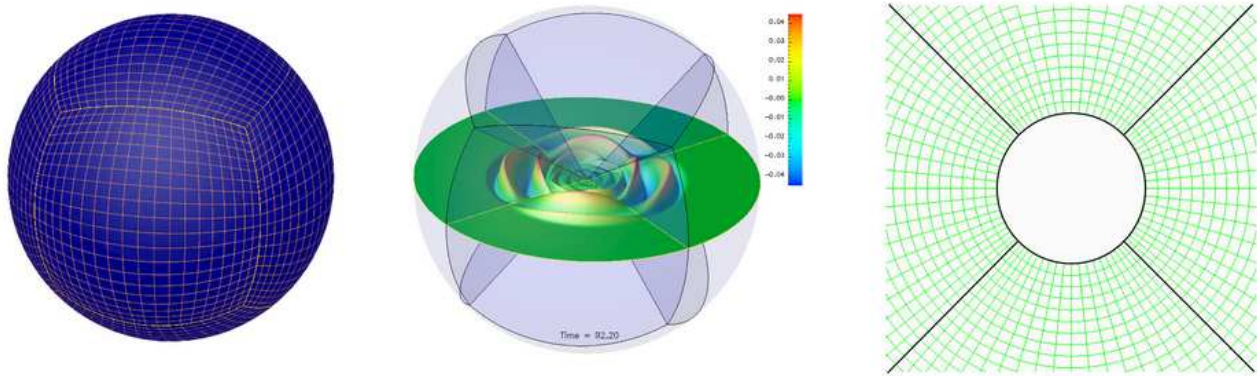} 
% \end{tabular}
\caption{Illustration of the six-block grid structure and the cubed sphere
  coordinates that are used for the simulations in this paper. The left panel
  shows the distribution of grid points on a sphere of constant radius. The
  central panel shows a snapshot from a scalar wave evolution on an equatorial
  cut.  The plot refers to an $\ell=m=2$ mode on the background of a Kerr
  black hole with spin $j=0.9$ at $t=92.2M$.  Also shown are the locations of
  the inter-block boundaries.  The right panel magnifies the central region of
  the domain in the equatorial plane, showing the grid structure around the
  spherical excision boundary.  The four dark lines mark the interfaces
  between blocks. }
  \label{fig:blocks}
  % \end{center}
\end{figure*}

We perform our evolutions describing scalar perturbations of a Kerr spacetime
through excision of the singularity. With our multi-block approach we can have
smooth (in particular, spherical) inner (excision) and outer boundaries.  As
in \cite{Lehner2005a}, we use a six-block setup with a global topology of
$S^2\times R^+$, referred to as \emph{cubed sphere} coordinates
(Fig.~\ref{fig:blocks}).  This topology and the corresponding coordinates on
each block are well adapted for modeling a single central object together
with outgoing radiation that is generated at or close to that object.

The six blocks are arranged like the six faces of a cube, i.e., block 0 covers
the neighborhood of positive $x$, block 1 positive $y$, block 2 negative $x$,
block 3 negative $y$, block 4 positive $z$, and block 5 negative $z$.  On each
of those blocks a local coordinate system $(\hat a,\hat b,\hat c)$ is defined,
with $-1 \le (\hat a,\hat b,\hat c) \le +1$, and equal grid spacing in the
local system.  The coordinate $\hat c$ runs along the radial direction, and
$\hat a$, $\hat b$ span the angular ones.  See \cite{Lehner2005a} for the
explicit definition of these coordinates.

\subsection{Background metric}

We consider a stationary, rotating black hole background.  The Kerr metric
can be written in Kerr-Schild form  as

\begin{equation}
\label{eq:kerrschildmetric}
ds^2 = \eta_{\mu \nu} + 2 H l_\mu l_\nu dx^\mu dx^\nu
\end{equation}
with $\eta_{\mu \nu}$ the Minkowski metric, and
\begin{eqnarray}
H &=& \frac{Mr}{r^2 + a^2 \left (z / r\right )^2}\,, \\
r^2 &=& {1 \over 2} (\rho^2-a^2) + 
\sqrt { {1\over 4} (\rho^2-a^2)^2 + a^2 z^2}\,, \\
\rho^2 &=& x^2 + y^2 + z^2\,.
\end{eqnarray}
Here $M$ is the mass and $a=jM=J/M$ is the angular momentum per unit mass of
the black hole ($j$ is the dimensionless spin parameter, $0\leq j\leq 1$).  In
Cartesian coordinates, the null vector $l_\mu$ is given by
\begin{equation}
l_\mu dx^\mu = dt + 
{ rx+ay \over r^2+a^2} dx +
{ry - ax \over r^2+a^2} dy +
{z \over r} dz\,.
\end{equation}

\begin{figure}[htbp]
  % \begin{center}
  \includegraphics[width=6cm]{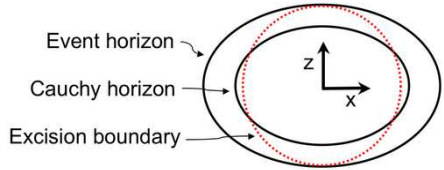}
  \caption{Event and Cauchy horizons for a Kerr black hole with spin
    $j\gtrsim 0.96$ in ``standard'' Kerr-Schild coordinates (as
    defined in the text), here shown in the $x$-$z$ plane.  
    The horizons have an
    ellipsoidal shape; it is therefore not possible to fit a spherical excision
    region (dotted line) between the two horizons.}
  \label{fig:horizons}
  % \end{center}
\end{figure}

This form of the Kerr-Schild metric has become of common use in numerical
relativity.  However, in these coordinates the shape of the Cauchy and event
horizons become more and more ellipsoidal with increasing spin\footnote{We
  thank Harald Pfeiffer for pointing this out to us.}.  For $j\gtrsim 0.96$ it
is not possible to fit a spherical excision boundary between these horizons
any more.  This is illustrated in Fig.~\ref{fig:horizons}.  Although we could
in principle choose a different shape for the excision boundary within our
code, we instead use coordinates in which both horizons are always spherical,
and therefore an excision sphere can always fit between them.  This version of
the Kerr Schild coordinates is related to the ``standard'' one defined above
by the following transformation:
\begin{subequations}
\begin{eqnarray}
\tilde x &=& x - {a y \over r}\,, \\
\tilde y &=& y + {a x \over r}\,, \\
\tilde z &=& z\,.
\end{eqnarray}
\end{subequations}

\subsection{Evolution system}

We write the time evolution equations for scalar perturbations in a symmetric
hyperbolic and (in the case of a stationary background) conservative form.
This guarantees stability and energy conservation for the continuum equations.
We use differencing operators that satisfy the summation by parts (SBP)
property, and this also guarantees stability and energy conservation in the
semi-discrete case (see e.g.\ \cite{Lehner04} for more details).  On a time
independent background the evolution equations take the form
\begin{eqnarray}
\label{eqn:wave equation}
\dot\Phi &=& \Pi\,, \\
\dot\Pi &=& \beta^i \partial_i \Pi + 
{\alpha \over \sqrt{h}} \partial_i 
\left( {\sqrt{h} \over \alpha} \beta^i \Pi + 
\alpha \sqrt{h} H^{ij} d_j \right)\,, \\
\dot d_i &=& \partial_i \Pi\,,
\end{eqnarray}
where $\Phi$ denotes the scalar field, $\Pi$ its time derivative, and
$d_{i}=\partial_i \Phi$ the spatial gradient of the field. The quantity
$h_{ij}$ is the three metric, $h$ its determinant, $h^{ij}$ the inverse three
metric, $\alpha$ the lapse, and $\beta^i$ the shift vector.  $H^{ij} =
h^{ij}-\beta^i \beta^j/\alpha^2$ is the spatial part of the inverse
four-metric.

The background geometry for all simulations presented here is that of
a Kerr black hole.  It would be possible to exploit the axisymmetry of
the background spacetime by performing a multipole decomposition of
the scalar field, and then solving for each azimuthal number $m$ as an
axisymmetric, two-dimensional problem.  We choose not to do so here
but instead solve the full three-dimensional equations.  This has the
advantage that we can later use the same implementation for generic,
non-axisymmetric spacetimes.  Using a fully three-dimensional code
also serves to test our numerical multi-block and excision techniques
in a scenario that is non-trivial, but at the same time not as
complicated as solving the full Einstein equations.

\subsection{Initial and boundary conditions} \label{initial}

The QNM excitation depends on the angular structure of the scalar
field that is used as a perturbation.  To excite certain modes in a
controlled way, we choose initial data of the form
\begin{subequations}
\begin{eqnarray}
\Phi &=& A \exp\left( -\frac{(r-r_0)^2}{\sigma^2} \right) Y_{\ell m} \label{eq:initialdata1}\,,\\
\Pi &=& B \exp\left( - \frac{(r-r_0)^2}{\sigma^2} \right) Y_{\ell m} \label{eq:initialdata2}\,,\\
d_{i} &=& \partial_i \Phi\,.
\label{eq:initialdata3}
\end{eqnarray}
\label{eq:initialdata}
\end{subequations}
Unless otherwise stated, throughout this paper we use $r_0=20M$ and 
$\sigma=M$.  $Y_{\ell m}(\theta, \phi)$ denotes the
ordinary spherical harmonics. Since the Kerr background is not
spherically symmetric, we should really expand the perturbation in
terms of spin-weighted spheroidal harmonics $_{s}S_{\ell m}(j\omega)$
of spin weight $s=0$. Using spherical harmonics weakly excites other
modes through rotational mode mixing; this point will be discussed in
more detail below, in Sec.~\ref{mixing}.

The changes in the characteristic length scale in the radial direction
are usually small over time.  To accurately resolve the propagating
waves all the way to the outer boundary we use a constant
 resolution in the radial direction of our cubed sphere coordinates. 
As mentioned, the coordinates are set up so that the spherical inner
(excision) boundary is placed between the event and Cauchy horizons,
and no boundary conditions need to be applied there.  For global
stability we choose maximally dissipative boundary conditions at the
outer boundary, and we apply them through penalty terms.  

%%%%%%%%%%%%%%%%%%%%%%%%%%%%%%%%%%%%%%%%%%%%%%%%%%%
\subsection{Specifications for the simulations} \label{specs}
%%%%%%%%%%%%%%%%%%%%%%%%%%%%%%%%%%%%%%%%%%%%%%%%%%% 

We use spatial finite differencing operators that satisfy summation by parts;
they are eighth order accurate in the interior and fourth order accurate at
and close to the boundaries.  With those operators we expect a global accuracy
of order five (see \cite{Diener05b} for more details on the operators that we
use).  We use a fourth order accurate Runge Kutta time integrator. This does
not spoil the expected global fifth order spatial convergence, since we use a
small enough time step so that the truncation errors generated by the time
integration are smaller than those that originate from the spatial finite
differencing (see \cite{Diener05b} for details on the code's convergence).

In multi-block simulations one does not necessarily have a uniform or
isotropic grid spacing in a global coordinate system. Since in all our
simulations the global grid spacing in the radial direction is smaller
than in the angular directions, we use the radial direction for our
time step criterion $\Delta t = \lambda \Delta r$, where $\lambda$
---usually referred to as the Courant factor--- is chosen to be
$\lambda=0.25 $.

Taking into account our initial data [cf.\ Eq.~(\ref{eq:initialdata})] and the
typical position of the observer, and given that we are interested in the
ringdown phase, evolution times of about $t=150M$ with outer boundaries at
about $200M$ are a reasonable choice.  Unless otherwise stated, for the
simulations that we show below we use ten points per length unit $M$ in the
radial direction and an angular resolution of $21 \times 21$ grid points per
block, which gives us approximately 80 grid points along the circumference of
any sphere with constant radius. As described below, we have found that with
this resolution we can get good agreement with the Kerr quasinormal
frequencies predicted by perturbation theory.

\begin{figure}[htbp]
\begin{center}
\includegraphics{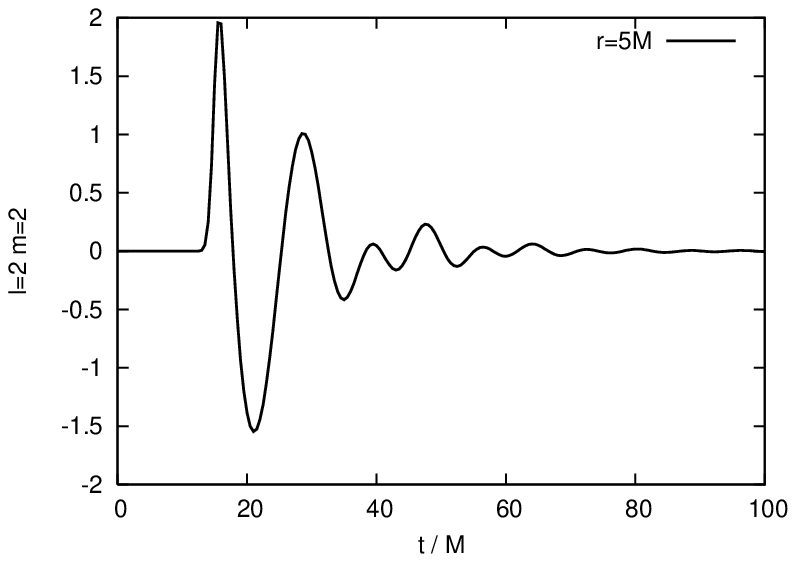}
\includegraphics{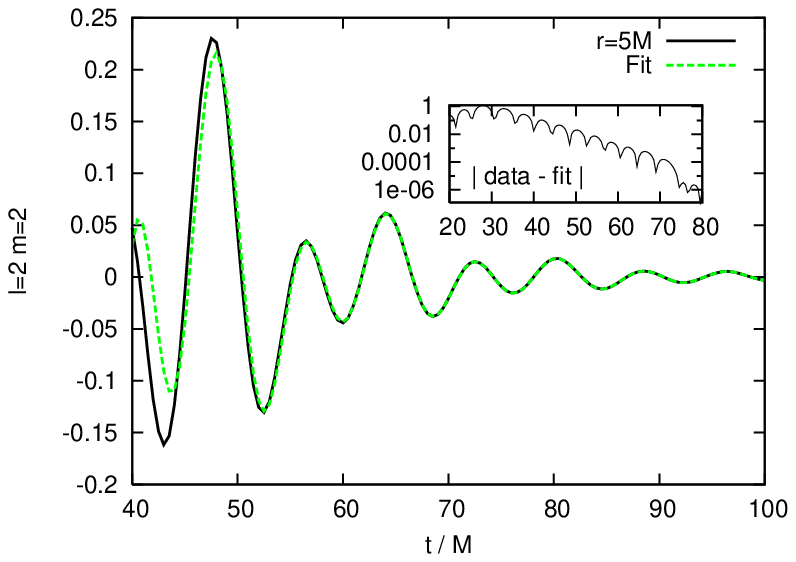}
\caption{The left panel shows the $\ell=m=2$ component of the waveform
  extracted at radius $r=5M$ on a Kerr black hole background with a spin of
  $j=0.9$. The waveform is a superposition of the corotating and the
  counterrotating mode, and the beating of two different frequencies is
  clearly visible. The right panel shows the waveform for $t\geq 40M$ as well
  as a QNM fit with the fundamental $\ell=|m|=2$ modes. The interval used for
  the fit is $[74.5M,150M]$. The inlay shows the absolute value of the
  difference between the fit and the data.  At times between the excitation of
  the QNM ($t\sim 25M$) and about $70M$ the differences are mainly due to the
  presence of the $(l=2, m=2, n=1)$ mode, the exponentially decaying mode that
  can be seen in the inlay (a fit of this mode yields quasinormal frequencies
  in agreement with perturbation theory). At times $t\lesssim 25M$ the
  difference is due to the initial burst.}
\label{fig:waveform}
\end{center}
\end{figure}

Figure \ref{fig:waveform} shows a typical waveform that we get when extracting
the real part of the $\ell=2, m=2$ mode from our simulations. The initial data
are set up according to Eq.~(\ref{eq:initialdata}), with the specific choice
$A=0$, $B=1$, $\sigma=M$ and $r_{0}=20M$. The background Kerr black hole has a
spin $j=0.9$. The strongest modes in this waveform are $(\ell=2, m=2, n=0)$
and $(\ell=2, m=-2, n=0)$, where we use $n$ to label overtones, $n=0$ being
the fundamental mode. We show a fit for those two modes together with the
numerical data. The third strongest component in the data is the $(\ell=2,
m=2, n=1)$ mode. Since this mode is decaying much faster than the fundamental
mode, it only plays a role at early times. That is the reason why our fit,
done for only the two fundamental modes, is drifting away from the numerical
data at times below $50M$ (we will explicitly analyze overtones in
Sec.~\ref{overtones}).

%%%%%%%%%%%%%%%%%%%%%%%%%%%%%%%%%%%%%%%%%%%%%%%%%%%
\section{Quasinormal mode frequencies and excitation amplitudes 
from numerical simulations}
\label{QNM}
%%%%%%%%%%%%%%%%%%%%%%%%%%%%%%%%%%%%%%%%%%%%%%%%%%%
\subsection{Overview}
The time evolution of perturbations of a Kerr black hole can be split
into three stages.  After a first burst of radiation depending on the
source of the excitation, the perturbation field $\Phi$ undergoes
exponentially damped oscillations (ringdown phase). Finally, in the
tail phase (caused by backscattering of radiation off the background
gravitational potential) the field follows a power-law decay. In this
paper we focus on the ringdown stage. We extract the different
multipole components of the numerical solution by integrating the
scalar field against different spherical harmonics over surfaces of
constant observer radius $r$:
\begin{equation}
\label{eq:modeextraction}
\Phi_{\ell m} (r,t)= 
\int Y_{\ell m}^*(\theta,\phi) \Phi\ d\Omega\,,
\end{equation}
where a star denotes complex conjugation. We usually consider
multipole components up to $\ell=4$ and all values of $m$ $(|m|\leq
\ell )$. By adding up the contributions of all multipoles one should
recover the full scalar field:
\begin{equation}
\label{eq:normrelation}
\int_{r} \Phi^2 d\Omega   = 
\sum_{\ell=0}^\infty \sum_{m=-\ell}^{\ell} (\Phi_{\ell m})^2\,.
\end{equation}
As already mentioned, the rotation of the black hole and numerical errors can
excite multipole components which are not present in the initial data.  The
above property can be used to check for the existence of overtones or modes
with $\ell>4$ that are not explicitly extracted but might be present in the
solution (e.g. due to rotational mode mixing, numerical errors, or both).
Multipoles with $m \neq 0$ require some care. The spherical harmonics $Y_{\ell
  m}(\theta, \phi)$ are given by
\begin{equation}
Y_{\ell m}(\theta, \phi) =
\sqrt{\frac{2\ell+1}{4\pi}\frac{(\ell-m)!}{(\ell+m)!}} P_{\ell}^m(\cos\theta)
 e^{\ii m \phi}\,,
\end{equation}
where $P_{\ell}^m(\theta)$ is a real function (an associated Legendre
polynomial). Therefore the initial data of a pure multipole with $m\neq 0$
will be complex.  Given that the evolution equations are linear, we can evolve
the real and imaginary parts of $\Phi$ separately, and obtain the complex
solutions for positive and negative $m$ by linear combinations of the form
\begin{subequations}
\begin{eqnarray}
\Phi_{\ell m} &=& \mathcal{R} (\Phi_{\ell m}) + 
\ii \mathcal{I} (\Phi_{\ell m})\,, \\
\Phi_{\ell -m} &=& \mathcal{R} (\Phi_{\ell m}) - 
\ii \mathcal{I} (\Phi_{\ell m})\,.
\end{eqnarray}
\end{subequations}
This point is important for the extraction of the relative amplitude of
corotating and counterrotating modes. In fact, as stressed (for example) in
Ref.~\cite{Berti06b,Leaver85}, QNMs of Kerr black holes always come ``in
pairs''.  In the Kerr case, for a given multipole $(\ell,m)$ we have to solve
an eigenvalue problem to determine both the quasinormal frequencies
$\omega_{\ell mn}$ and the angular separation constant $A_{\ell mn}$ (not to
be confused with the mode amplitude ${\cal A}_{\ell mn}$ introduced below),
used to separate the angular and radial dependence of the Teukolsky equation
and write it as two ordinary differential equations.  For {\em each} $(\ell,m
\ne 0)$ and $j\neq 0$ the eigenvalue problem admits {\it two} sets of
solutions. In addition to $(\ell,m)$, we label the modes of each set by the
overtone index $n$, denoting the frequencies by $\omega_{\ell mn}^{(i)}$
$(i=1,2)$.  For given $(\ell,m,n)$, the solutions corresponding to the two
different sets have different values of $\omega_{\ell mn}$ (and also of
$A_{\ell mn}$):
$$
\omega_{\ell mn}^{(1)} \neq \omega_{\ell mn}^{(2)}\, .
$$
Both the real and imaginary parts are different.  In fact, the real part of
one of the frequencies is positive and the other one is negative:
$$
\mathcal{R}(\omega_{\ell mn}^{(1)})>0 
\, , \qquad
\mathcal{R}(\omega_{\ell mn}^{(2)})<0 \, .
$$
If we consider instead the frequencies corresponding to the pair $(\ell, -m)$,
they are related to those of $(\ell, m)$ by a simple symmetry property:
\be\label{minusm} -\mathcal{R}(\omega_{\ell
  mn}^{(i)})=\mathcal{R}(\omega_{\ell-mn}^{(j)})\,, \qquad
\mathcal{I}(\omega_{\ell mn}^{(i)})=\mathcal{I}(\omega_{\ell-mn}^{(j)})\,,
\qquad \left(A_{\ell mn}^{(i)} \right)^*= A_{\ell-mn}^{(j)}\,, \;\;\;
(i,j=1,2\,; i \neq j)\,. \ee
In this sense, any solution with positive $m$ is nothing but the ``mirror
image'' of a solution with opposite real part and opposite $m$ (see Fig.~6 of
\cite{Berti06b} for an illustration of this).  For $m=0$ (or for any value of
$m$ in the Schwarzschild case) the two ``mirror solutions'' are degenerate in
modulus of the frequency and damping time. However, in general, a multipolar
component with a given $(\ell,m)$ will always contain a superposition of {\it
  at least} two different damped exponentials. Because of this, it is enough
to consider only one frequency for each mode [$(\ell,m)$ or $(\ell, -m)$],
since the other two frequencies are obtained through this symmetry property;
we follow the standard convention of considering, for each mode, the frequency
with positive real part. Below we will discuss in detail the excitation of
these modes, extending previous work by Krivan {\it et al.} \cite{Krivan97a}.

When the perturbation field is in the quasinormal ringing regime, it
can be expanded as a QNM sum of the form
\be
\Phi_{\ell m}(r,t)\approx 
\mathcal{R}\left\{
\sum_{n=0}^\infty {\cal A}_{\ell m n} e^{\ii c_{\ell m n}}e^{-\ii
  \omega_{\ell m n}(t-t_0)} 
\label{expansion}
\right\}\,,
\ee
where ${\cal A}_{\ell mn}$ is the amplitude of the $n$-th overtone with
angular structure given by the pair $(\ell,m)$, $c_{\ell mn}$ its phase,
$\omega_{\ell mn}$ its complex quasinormal frequency and $t_0$ (which to a
first approximation we assume to be the same for all modes) marks the time at
which the quasinormal regime starts.

The extraction of gravitational waves from numerical simulations of the full
Einstein equations requires the observer to be located far away (in the wave
zone). For the extraction of QNM frequencies, on the other hand, it is not
problematic to place the observer close to the black hole, since an observer
at any point in the space time is in general expected to measure the same
frequencies. In fact, a small $r$ is better suited for extracting
quasinormal frequencies from our simulations simply because outer boundary
effects pollute our waveform later, and the ringing regime can be observed for
a longer time.  The availability of a longer ringdown waveform improves the
accuracy of the fitting procedure that we apply to extract the frequencies.

%%% This Table contains values for  a resolution of M/10
\begin{table}[htbp]
   \centering
   \caption{Quasinormal frequencies computed by Leaver's continued
     fraction method (here labeled ``perturb.'') and by our time domain
     simulations, with the associated relative differences. We use $21\times21$
     points in the angular direction on each block and a resolution of
     $M/10$ in the radial direction. For $j=0.9$ we compare the
     frequencies as seen by observers located at different radii $r$. 
     Observers at larger radii measure frequencies with
     larger errors, since boundary effects start to contaminate the waveform 
     earlier.}
  \begin{tabular}{|l|l|l|l|l|l| }
  \hline
  $r$ & $j$ & $l$, $m$ & $\omega_{\mathrm{perturb.}}$  & $\omega_{\mathrm{numerical}}$  & rel.\ difference (Re,Im) \\ 
    \hline \hline
     $5M$ & $0.0$ & $2$, $0$ & $0.48364-0.09676\ii$ & $0.48364-0.09676\ii$  & $<10^{-5}$ \\ 
     & $0.5$ & $2$, $0$ & $0.49196-0.09463\ii$ & $0.49190-0.09469\ii$ & $4.27 \times 10^{-4},  6.34 \times 10^{-4}$ \\ 
   &   $0.5$ & $2$, $-2$ & $0.42275-0.09562\ii$ &  $0.42281-0.09569\ii$ & $1.42 \times 10^{-4}, 7.32 \times 10^{-4}  $ \\ 
    & $0.5$ & $2$, $2$ & $0.58599-0.09349\ii$ & $0.58589-0.09339\ii$ & $1.71 \times 10^{-4},  1.07 \times 10^{-3}$ \\
   &  $0.9$ & $2$, $0$ & $0.51478-0.08641\ii$ & $0.51471-0.08646\ii$ &  $1.36 \times 10^{-4}, 5.79 \times 10^{-4}$ \\ 
   &  $0.9$ &   $2$, $-2$ & $0.38780-0.09379\ii$ & $0.38781-0.09339\ii$ & $2.58 \times 10^{-5}, 4.26 \times 10^{-3}$ \\ 
    & $0.9$ & $2$, $2$ & $0.78164-0.06929\ii$ & $0.78144-0.06955\ii$ & $2.56 \times 10^{-4}, 3.75 \times 10^{-3}$ \\ 
    & $0.98$ & $2$, $2$ & $0.89802-0.04090\ii$ & $0.90940-0.04018\ii$ & $1.27 \times 10^{-2}, 1.76 \times 10^{-2}$ \\
     & $0.98$ & $2$, $-2$ & $0.38177-0.09338\ii$ & $0.38234-0.09743\ii$ & $1.49 \times 10^{-3}, 4.34 \times 10^{-2}$ \\
\hline
$20M$ & $0.9$ & $2$, $-2$ &   $0.38780-0.09379\ii$ & $0.38694-0.09471\ii$ &     $2.22\times 10^{-3}$, $9.81\times 10^{-3}$ \\
 & $0.9$ & $2$, $2$ & $0.78164-0.06929\ii$ &      $0.78244-0.06670\ii$ &    $1.02\times 10^{-3}$, $3.74 \times 10^{-2}$ \\
\hline
$40M$ & $0.9$ & $2$, $-2$ &   $0.38780-0.09379\ii$ &       $0.38406-0.09958\ii$ &   $9.64\times 10^{-3}$, $6.17 \times 10^{-2}$ \\
& $0.9$ & $2$, $2$ & $0.78164-0.06929\ii$ &       $0.78292-0.06618\ii$ &  $1.64\times 10^{-3}$,  $4.49 \times 10^{-2}$ \\

\hline
  \end{tabular}
   \label{tab:freq}
\end{table}

The effect of the observer's location on the result is illustrated in Table
\ref{tab:freq}, where we list the frequencies of $(\ell=2,m=\pm2)$ fundamental
modes for a Kerr black hole with spin $j=0.9$ as measured by observers at
radii $r=5M$, $20M$ and $40M$. We picked $t_{0}=r+r_0$ in
Eq.~(\ref{expansion}) and $A=0$, $B=1$ in Eq.~(\ref{eq:initialdata}). The
results presented in this Table are discussed in more detail below
(Sec.~\ref{sec:fittingprocedure}). Here we simply remark that quasinormal
frequencies measured at different radii are very close to the analytical
predictions, supporting the statement that the observer does not need to be
far away from the black hole to extract the correct ringdown frequencies.
Indeed, {\em for these particular simulations} the relative error increases
with $r$: the main reason, as explained, is that observers located
at large radii see boundary effects earlier, so they can only measure a
shorter ringdown waveform with respect to observers closer to the black hole.

\subsection{The time shift problem} \label{timeshiftproblem}

Here we discuss the so-called {\it time-shift problem}, how it affects the
extraction of quasinormal frequencies and amplitudes from numerical
simulations, and a possible way to address it.  Even though in this paper we
consider scalar perturbations, the discussions of this and other sections
apply also to other types of black hole perturbations.

The standard approach is to choose $t_0$ in Eq.~(\ref{expansion}) using some
approximate calculation based, for example, on the location of the initial
data and the time it would take for initial data to be scattered by the black
hole potential and reach the observer, usually assuming that perturbations
propagate with coordinate speed one (as they would in flat spacetime).
Criteria like this are well motivated and provide a good guess, but there is
still an uncertainty in $t_0$. For example, the coordinate speed of the
perturbation in a curved background in general will not be one. One might
expect that such a small uncertainty would not influence the extraction of
physically relevant quantities. However, as we discuss below, this is not the
case: there are quantities of interest to gravitational wave detection which
have a strong dependence on $t_0$. Following the existing literature, we will
call this the {\it time-shift problem}.

Suppose the starting time $t_0$ is subject to an uncertainty
$\delta_0$. Under a change
\be
t_0 \rightarrow t_0 + \delta_0 \label{delta0}\,,
\ee
the amplitude and phase of each mode change according to 
\begin{subequations}
\begin{eqnarray}
{\cal A}_{\ell mn} &  \rightarrow & {\cal A}_{\ell mn}'=
{\cal A}_{\ell mn}
e^{-\delta_0\mathcal{I}\left(\omega_{\ell mn}\right)}\,, \label
{amplitude_error}\\
c_{\ell mn} & \rightarrow & c_{\ell mn}' =
c_{\ell mn} + 
\delta_0 \mathcal{R}\left( \omega_{\ell mn} \right)\,.
\end{eqnarray}
\end{subequations}
That is, an uncertainty in $t_0$ induces a linear uncertainty in the
phase, and an {\em exponential} uncertainty in the
amplitude. Fortunately other quantities are largely independent of
this uncertainty: for example, the QNM
frequencies $\omega_{\ell mn}$ are unaffected by
$\delta_0$.

How large can we allow this exponential amplification of errors to be?
Let us require the amplitude uncertainty induced by the starting-time
uncertainty $\delta_0$ to be less than some small number $\epsilon$,
that is
$$
\left|\frac{{\cal A}_{\ell mn}'-{\cal A}_{\ell mn}}
{{\cal A}_{\ell mn}}\right|=
\left|e^{-\delta_0\mathcal{I}\left(\omega_{\ell mn}\right)}-1\right|
<\epsilon\,.
$$
For small $\epsilon$ this implies
\be
|\delta_0|\lesssim
\left|
\frac{\epsilon}{\mathcal{I}\left(M\omega_{\ell mn}\right)}
\right|M\,.
\ee
For the $\ell=2$ fundamental scalar mode in the Schwarzschild
background (which is spherically symmetric, so that the choice of $m$
becomes irrelevant) $|\mathcal{I}\left(M\omega_{200}\right)|=0.09676\simeq
10^{-1}$. In other words, if we want to determine the amplitude of
this mode within $1\%$ ($\epsilon=10^{-2}$) we need to know $t_0$ with
an uncertainty $\delta_0\lesssim 0.1M$. Constraints on $\delta_0$ are
even tighter for overtones, since they decay faster and the
exponential propagation of errors is more dramatic.

In practice, what is most interesting is the {\em relative} amplitude
between different modes. Under a change of the form (\ref{delta0})
this relative amplitude changes according to
\be
\label{eq:timeshiftrelamp}
\frac{{\cal A}_{\ell mn}}{{\cal A}_{\ell' m'n'}} \rightarrow 
\left(\frac{{\cal A}_{\ell mn}}{{\cal A}_{\ell'm'n'}}\right)'=
\frac{{\cal A}_{\ell mn}}{{\cal A}_{\ell'm'n'}}
e^{-\delta_0\mathcal{I}\left(\omega_{\ell mn}-\omega_{\ell'm'n'}\right)}\,.
\ee
Following the same reasoning we find the constraint
\be\label{delta0crit}
|\delta_0|\lesssim
\left|
\frac{\epsilon}{M\mathcal{I}\left(\omega_{\ell mn}-\omega_{\ell'm'n'}\right)}
\right|M\,.
\ee
Consider for example the relative amplitude between the fundamental
mode and the first overtone. For Schwarzschild black holes and small
values of $n$ the typical difference in the imaginary part of the
frequency for two consecutive overtones ($\ell'=\ell$, $m'=m$,
$n'=n+1$) is
$$
M\mathcal{I} \left(\omega_{\ell mn}-\omega_{\ell'm'n'} \right)\simeq 0.2\,.
$$ 
Setting again $\epsilon=10^{-2}$ the maximum allowed uncertainty on
the starting time would be quite small: $\delta_0\lesssim 0.05 M$
(this presumably already precludes assuming that the perturbation
propagates with speed one, as in flat spacetime). 

Suppose we want to resolve corotating and counterrotating components
of the fundamental mode with $\ell=2$ (say, the components with $m=\pm
\ell$). In the case of a spinning black hole background these QNM
frequencies are different, but their imaginary parts are actually
quite close for most values of the rotation rate
\cite{Leaver85,Berti06b}.
For example, looking at Table \ref{tab:freq} we see that for spin
$j=0.5$ the difference is $|M\mathcal{I} \left(\omega_{220} -
\omega_{2-20} \right)| \simeq 0.00212$, so that $\delta_0\lesssim
4.7M$.  Even for a rapidly rotating black hole with $j=0.9$ the
difference is not as large as between a fundamental mode and its
overtone: $M\mathcal{I} \left(\omega_{220} - \omega_{2-20} \right)
\simeq 0.0245$, and $\delta_0\lesssim 0.4M$.

\begin{figure}[htbp]
\begin{center}
\begin{tabular}{cc}
\includegraphics{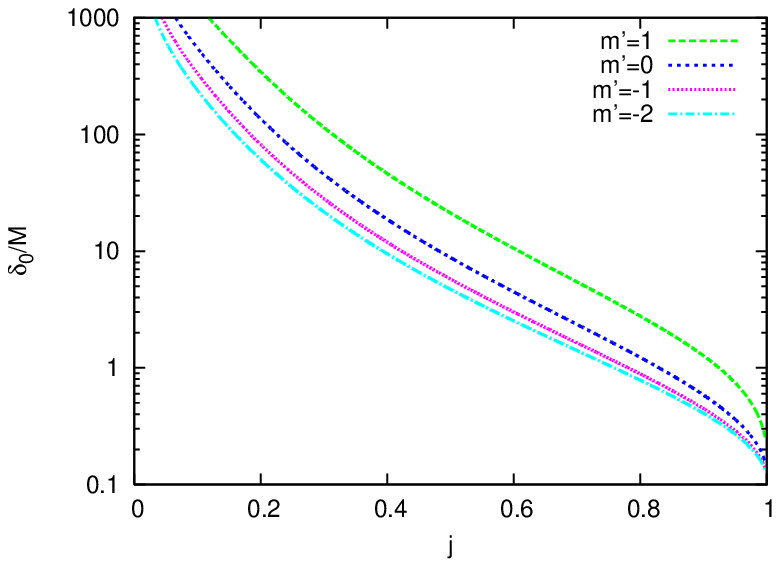}&
\includegraphics{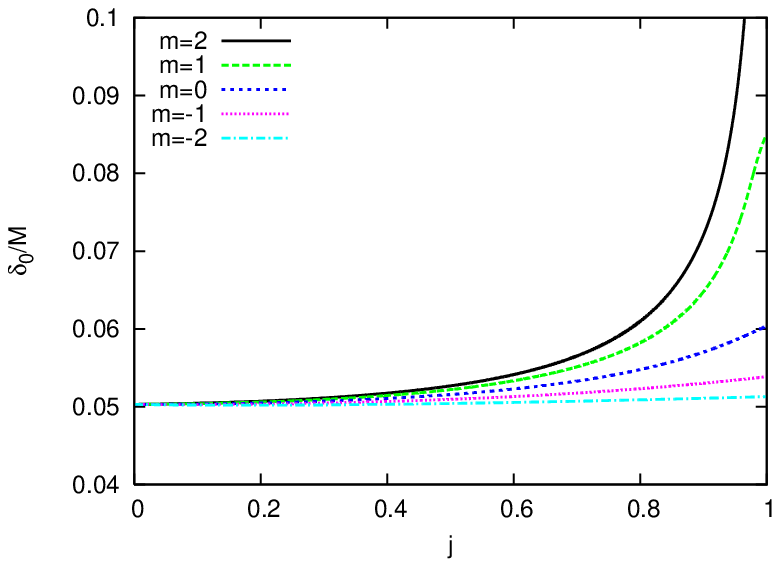} \\
\end{tabular}
\caption{Critical uncertainty in the starting time, as defined by
Eq.~(\ref{delta0crit}), assuming $\epsilon=10^{-2}$. In the left panel
we give the critical $\delta_0$ for fundamental modes ($n=n'=0$) with
different angular dependence. For the first mode we assume $\ell=m=2$;
the second mode has $\ell '=2$ and different values of $m'=1,0,-1,-2$
(lines from top to bottom). In the right panel we show the critical
uncertainty in the relative amplitude of the fundamental mode and
first overtone, i.e., $n=0$ and $n'=1$. Here we set $\ell = \ell '=2$,
consider all values of $m=m'$ and once again we assume
$\epsilon=10^{-2}$. }
\label{fig:delta0}
\end{center}
\end{figure}

Critical starting-time uncertainties for $\epsilon=10^{-2}$, general values of
the spin and different pairs of modes are plotted in Fig.~\ref{fig:delta0}.
Determining the relative amplitude of a fundamental mode and of the first
overtone is generally harder, unless we consider corotating modes and
near-extremal black holes, as we do in Sec.~\ref{overtones}. The spin
dependence of $\delta_0$ is quite weak for overtones, but $\delta_0$ can
change by orders of magnitude for modes with different angular dependence
($\ell\neq \ell'$ or $m\neq m'$).  For $j\lesssim 0.5$ the {\it time-shift problem},
as we defined it here, becomes irrelevant when we want to determine the
relative amplitude of components with the same $l$ and different $m$'s. The
reason is simply that modes with different $m$'s have the same QNM frequency
in the Schwarzschild limit, so that $\delta_0\to \infty$. As a rule of thumb,
determining the relative amplitude of angular components with the same $l$ and
different $m$'s is harder for large rotation. However, as we said before, even
for $j=0.9$ the critical uncertainty is $\delta_0\gtrsim 0.4M$, an order of
magnitude larger than the typical uncertainty to resolve overtones (which in
most cases is $\sim 0.05M$). Most of the qualitative features of
Fig.~\ref{fig:delta0} are also seen in the experimental problem of resolving
different QNMs in the actual detection of a ringdown signal (compare e.g.
Figs.~3, 4 and 18 of \cite{Berti06b}).

In Sec.~\ref{sec:corot} and Sec.~\ref{overtones} we will study in more detail
the extraction of corotating and counterrotating modes and of overtones,
respectively.  In preparation for this study, in the next Section we outline
the general method by which we extract quasinormal frequencies from our
numerical waveforms.

%%%%%%%%%%%%%%%%%%%%%%%%%%%%%%%%%%%%%%%%%%%%%%%%%%%
\subsection{Extraction of QNM frequencies by an optimal choice of the fitting interval}
%%%%%%%%%%%%%%%%%%%%%%%%%%%%%%%%%%%%%%%%%%%%%%%%%%%
\label{sec:fittingprocedure}

Once we have the different multipole components of the numerical solution, we
analyze them by applying a fitting procedure to each of these components.
Since each mode decays exponentially while oscillating with its quasinormal
frequency, the obvious function to fit the numerical waveform is
Eq.~(\ref{expansion}), where the free parameters are the amplitudes, phases
and frequencies. As discussed in Sec.~\ref{overtones}, only in some cases we
have been able to fit for overtones, in the sense of getting their expected
quasinormal frequencies with reasonable accuracy. However, as described below,
the residual that we get by truncating the sum at the fundamental mode is
already quite small (see also Fig.~\ref{fig:waveform}).

In this subsection we are interested in extracting the quasinormal frequencies
from our numerical data. To a very good approximation the frequencies are
independent of $t_0$, and we can therefore pick any value for the latter. We
still need to find a good choice for the time interval $[T_{i}, T_{f}]$ over
which the ringdown dominates and the fitting procedure works best. Since in
principle the parameters obtained from the fitting might depend on the choice
of this time interval, we discuss our procedure in detail.

Only during the ringdown phase does the waveform have the functional behavior
of Eq.~(\ref{expansion}), so the time interval $[T_{i}, T_{f}]$ should not
include the transient regime and the tail phase.  For our simulations we found
it reasonable to pick $T_{f}=150M$, since for $T>T_{f}$ the system typically
goes into the tail phase. The choice of $T_{i}$ is more delicate: small values
would bring the fitting time window out of the ringdown phase, but large
values would make the fitting interval small and the resulting fit inaccurate.
We decided to take a pragmatic approach: for different values of $T_i$ we
compute the (relative) residual $R(T_i,t_0)$ between the fitted function and
the numerical data, which we define as
\begin{equation}
        R(T_{i},t_0) = \left( \sum_{t_j=T_{i}}^{T_f} | \Phi_{\mathrm{data}}(t_j) - \Phi_{\mathrm{fit}}(t_j, t_0) | \right)
                \left(\sum_{t_j=T_i}^{T_f} | \Phi_{\mathrm{data}}(t_j)| \right)^{-1}
\label{def-res}
\end{equation}
We then choose the value of $T_{i}$ that minimizes the residual. In a
very well defined sense, this gives an optimal choice for $T_i$.  In
principle one could use other norms (for example, a sum over squares
instead of a sum over absolute values), but we checked that this does
not affect significantly the results of this paper. Choosing the value of $T_i$
that minimizes the residual defined above should not be confused with
the minimization procedure done at each $T_i$ to get the fit itself. 

Instead of extracting the quasinormal frequencies through a fitting
procedure, in principle one could also perform a Fourier transform of
the solution, as in Ref.~\cite{Krivan97a}. However we have found that
the fitting procedure provides us with far superior accuracy, even in
cases with relatively few sampling points. Nonetheless we compared to
the results that we obtained by Fourier analysis and found consistency between
both methods.

\begin{figure}[htbp]
\begin{center}
\includegraphics{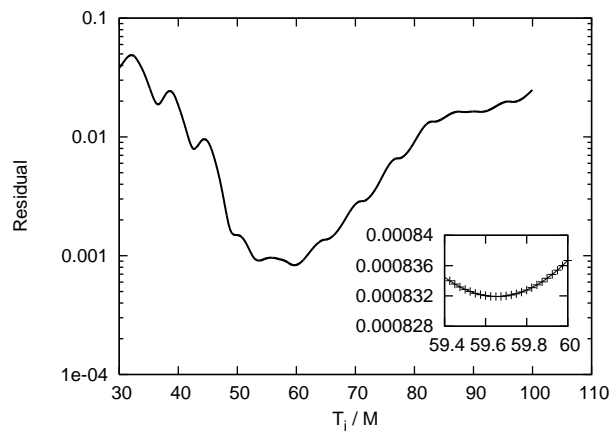}
\caption{Residual in the fit, as defined in Eq.~(\ref{def-res}), as a function
  of the initial time for the fitting $T_i$. Looking at the minimum of the
  residual we can determine $T_i$ with high precision. This plot corresponds
  to a simulation with spin $j=0.5$, $\ell=m=2$ initial data with $A=0$, $B=1$
  and a radial resolution $\Delta r = M/10$.}
\label{fig:residual}
\end{center}
\end{figure}

Figure \ref{fig:residual} shows the residual as a function of $T_i$ for one of
our simulations (the one corresponding to spin $j=0.5$ and $\ell=m=2$ initial
data in Table \ref{tab:freq}).  The residual is independent of the choice of
excitation time $t_0$, since a change in $t_{0}$ is just absorbed in the
amplitude of the fitting function, leaving the other fitting quantities
unaffected. 

Since the black hole's spin is non zero, both $m=2$ and $m=-2$ modes are
present in the solution. Here we discuss only the $m=2$ part of the numerical
solution. The $m=-2$ part behaves similarly (in Sec.~\ref{sec:corot} we
present a detailed study of the relative amplitudes of corotating and
counterrotating modes).

From Fig.~\ref{fig:residual} we see that $R(T_{i},t_0)$ has a rather sharp
local (and global) minimum. By computing the derivative (through
finite differences) of the residual with respect to $T_i$ we find that
the minimum is located at $T_i=(59.65 \pm 0.025)M$. The uncertainty
refers to the difference between two consecutive values of $T_i$,
which is in turn given by the time step for this simulation: $\Delta t =
0.025M$. 

\begin{figure}[htbp]
\begin{center}
\includegraphics{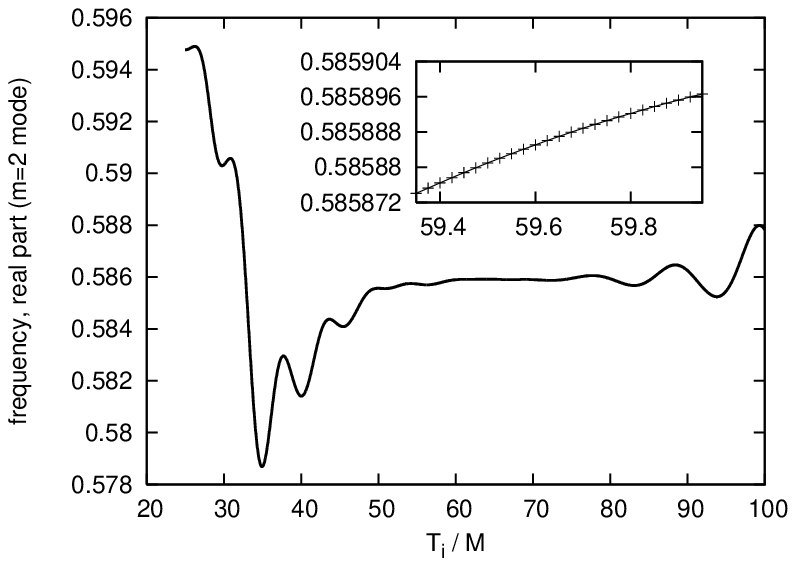}
\includegraphics{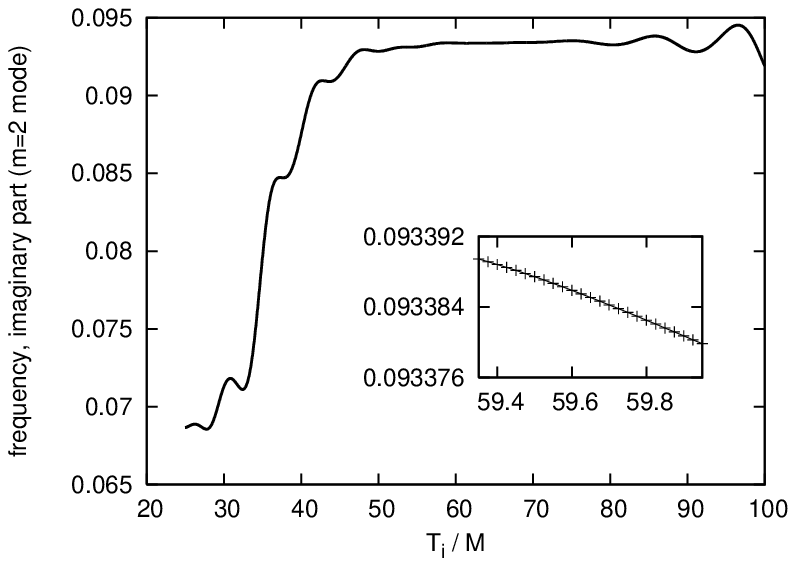}
\caption{The left and right panels show the real and imaginary parts
of the quasinormal frequencies extracted from the simulations in
Fig.~\ref{fig:residual}. From the optimal starting time determined by
minimizing the residual, $T_i=(59.65 \pm 0.025)M$ (see previous figure), 
we find $\omega_R = 0.585887 \pm 1 \times 10^{-6}$ and $\omega_I =
0.0933851 \pm 5 \times 10^{-7}$.}
\label{fig:residual2}
\end{center}
\end{figure}

Figure \ref{fig:residual2} shows the real and imaginary parts of the frequency
extracted from the same simulation as a function of $T_i$. By evaluating them
at $T_i=(59.65 \pm 0.025)M$ we get $\omega_R = 0.585887 \pm 1 \times 10^{-6}$
and $\omega_I = 0.0933851 \pm 5 \times 10^{-7}$.

Figure \ref{fig:residual2} also reveals that $\omega$ changes very little
within the interval $50M\lesssim T_i\lesssim 80M$. Since our choice of $T_i$
is by no means unique ---for example a different definition of the residual
would slightly shift $T_i$--- this plateau in the frequencies guarantees that
the physical quantities we extract are not too sensitive to that uncertainty.
This means that the errors in our numerically extracted QNM frequencies due to the
choice of $T_i$ are quite small.

We are now ready to examine the quasinormal frequencies obtained from our
numerical data in the way just described.  Table \ref{tab:freq} shows the
frequencies computed in \cite{Berti:2006wq} using Leaver's continued fraction
method for perturbed Kerr black holes with spin $j=0$, $0.5$, and $0.9$ (here
labeled {\it perturb.}). Along with these frequencies we list values extracted
from our time domain evolutions (labeled {\it numerical}) and the relative
differences between the two. The numerical values were obtained by evolving
different initial data sets with $A=0$ and $(\ell=2,m=0,~\pm 2)$ in
Eq.~(\ref{eq:initialdata}), and fitting for the multipoles present in the
initial data (we discuss the additional multipoles generated by rotational
mode mixing below).  For $j=0$ the frequencies do not depend on $m$, therefore
we only show results for $m=0$.  Even with a relatively modest resolution, the
differences on quasinormal frequencies from our three-dimensional simulations
in Table \ref{tab:freq} are between one and two orders of magnitude smaller
than the ones reported in previous two-dimensional, axisymmetric simulations
of gravitational perturbations \cite{Krivan97a}.

\subsection{Rotational mode mixing} \label{mixing}

In Sec.~\ref{initial} we described our initial data family sets, which were
expanded in spherical harmonics.  Since the Kerr background is not spherically
symmetric we should not expand the perturbation in terms of spherical
harmonics, but (more rigorously) in terms of the spin-weighted spheroidal
harmonics $_{s}S_{\ell m}(a\omega)$, where $s$ is the spin weight of the
perturbing field, $a=jM$ is the black hole spin parameter, and $\omega$ is the
frequency in a Fourier expansion of the perturbation (a quantitative
discussion of spin-weighted spheroidal harmonics and more references can be
found in \cite{Berti06c}).  However, as first shown by Press and Teukolsky
\cite{Press73}, the $_{s}S_{\ell m}$'s may be expanded as a power series in
$a\omega$:
\be
_{s}S_{\ell m} =~_{s}Y_{\ell m} + (a\omega)\,
\sum_{\ell^\prime \ne \ell} c_{\ell^\prime \ell m}~_{s}Y_{\ell^\prime m} 
+ O(a\omega)^2 \,.
\label{sexpand}
\ee 
Here $_{s}Y_{\ell m}$ denotes a spin-weighted spherical harmonic of
spin-weight $s$.  In this paper we focus on scalar perturbations ($s=0$), in
which case the spin-weighted spherical harmonics reduce to ordinary spherical
harmonics.  The coefficients $c_{\ell^\prime \ell m}$ are related to the more
familiar Clebsch-Gordan coefficients \cite{Press73,Berti06c}.  As a result of
(\ref{sexpand}), and because of the orthogonality of the (spin-weighted)
spherical harmonics, inner products of different spheroidal harmonics will be
given by inner products of \emph{spherical} harmonics with higher-order
corrections in $a\omega$.  At least for small $a\omega$, we may expect these
contributions to be small.  In fact, the corrections turn out to be small even
for moderately large values of $a\omega$ (see \cite{Berti06c} for an explicit
calculation of the inner products at the QNM frequencies).  Nevertheless,
using spherical harmonics instead of spheroidal harmonics can induce a small
amount of mode-mixing in the initial data.

For a spherically symmetric background spacetime, initial data with different
values of $\ell$ evolve separately and the angular structure of each
mode is preserved during evolution.  On the other hand, for a Kerr
background with nonzero spin, modes with different values of $\ell$ do
couple and furthermore, modes that are not present in the initial
data can be excited during evolution.  This may make it necessary to
increase the angular resolution compared to the non-rotating case to
resolve the higher $\ell$ modes generated during evolution.  However,
the decay rate of these modes increases with $\ell$, so even when
modes with higher values of $\ell$ are generated during evolution,
they do not dominate.  Therefore, we found that if we accurately resolve the angular
structure initially, the same is in general true for the whole evolution.

\begin{figure}[htbp]
\begin{center}
\includegraphics{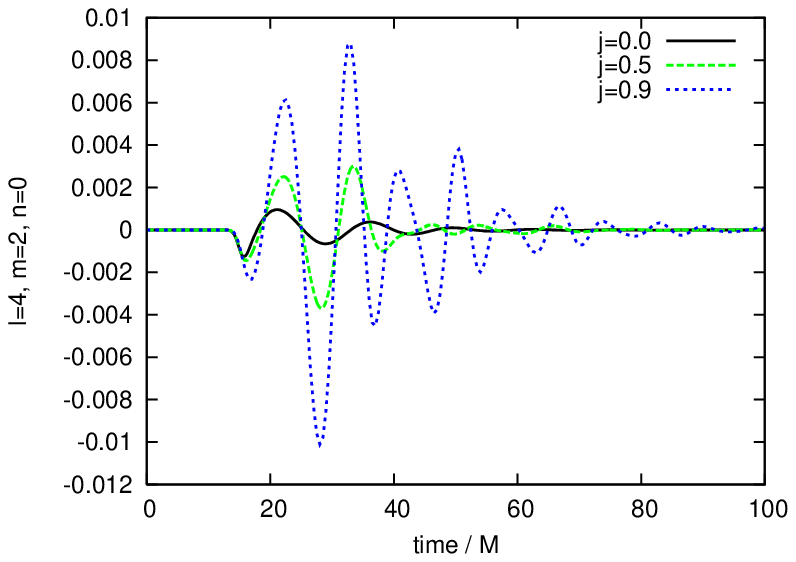}
\includegraphics{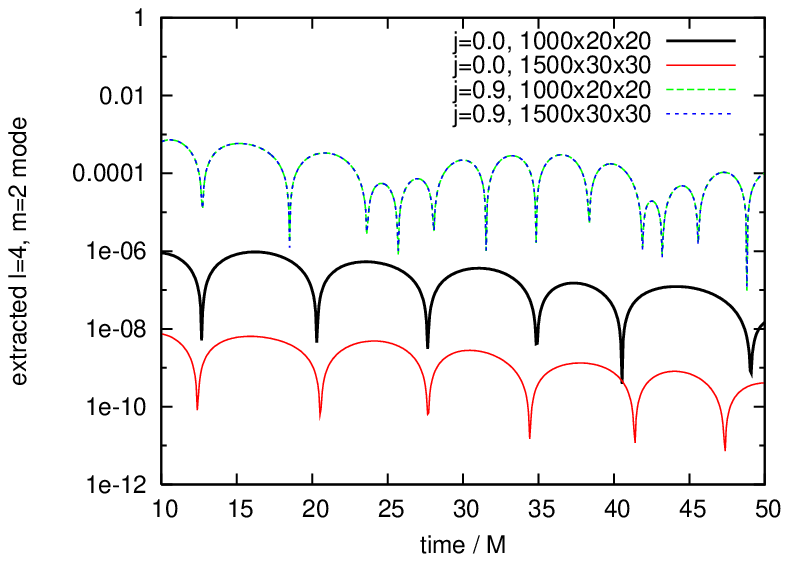}
\caption{The left panel shows the extracted $(\ell=4,m=2,n=0)$ waveform for three
  simulations with different spin parameters as seen by an observer at
  $r=5M$.  The initial data are a pure $(\ell=m=2)$ mode and
  are set up according to Eq.~(\ref{eq:initialdata}) with $A=0$, $B=1$ and
  $r_0=20M$.  For zero spin the different multipole components of the solution
  should evolve independently and no modes besides the one in the initial data
  should be excited, while for non-zero spin modes with different $\ell$ but
  same $m$ do couple \cite{Krivan97a}.  In the Schwarzschild case the
  $(\ell=4,m=2,n=0)$ waveform differs from zero due to our grid structure and
  discretization errors, but it converges to zero with increasing resolution.
  This is illustrated by the right panel, which shows the extracted
  $(\ell=4,m=2,n=0)$ amplitude for $j=0$ and $j=0.9$ from runs with two
  resolutions ($20\times 20 \times 1000$ and $30 \times 30 \times 1500$ points
  per patch and outer boundaries at $100M$). Only for $j=0.0$ the mode
  converges to zero. }
\label{fig:modemixing}
\end{center}
\end{figure}

Figure \ref{fig:modemixing} illustrates rotational mode coupling for non-zero
spin backgrounds (see also \cite{Zlochower03} and \cite{Allen97a} for
numerical studies of mode-mode coupling). Since modes with same $m$ but
different $\ell$ can couple to each other, we show the extracted
$(\ell=4,m=2,n=0)$ waveform (for three simulations with different spin
parameters) excited by initial data whose angular dependence is given by an
$\ell=m=2$ spherical harmonic.  As expected, the rotationally-induced
excitation of the $(\ell=4, m=2)$ mode typically increases with spin. Some
additional mode mixing is an artifact of the symmetry of our computational
grid. This ``spurious'' mode mixing is present also for $j=0$, but it
converges to zero as we increase the angular resolution. All other modes we
extract, up to $\ell=4$ and all allowed values of $m$, are within roundoff
error throughout the simulations.

\begin{figure}[htbp]
\begin{center}
\includegraphics[width=3in]{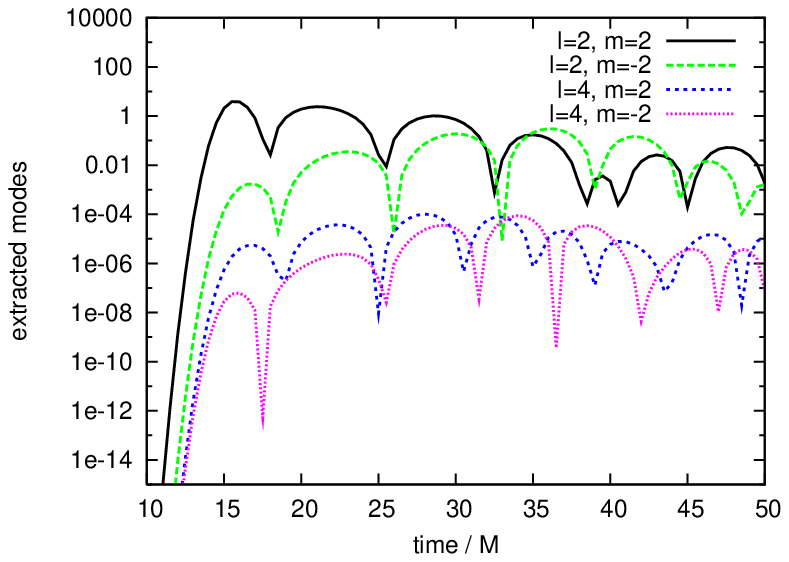}
\includegraphics[width=3in]{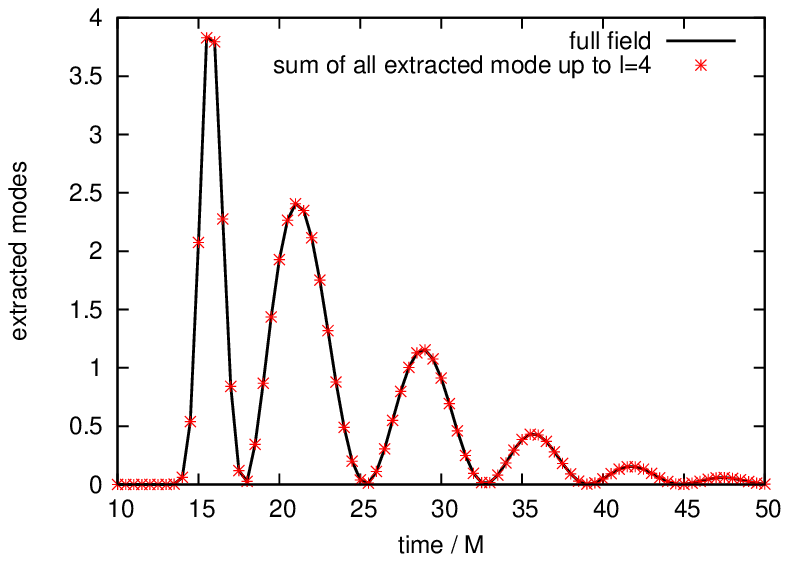}
\caption{Results from a run with initial data parameters $\ell=m=2$ and spin
  $j=0.9$. The left panel shows the square of the amplitude of all modes up to $\ell=4$
  which are not within the roundoff error. The right panel shows the sum over
  the square of those modes compared to the integral over a sphere of the full field squared: 
  As expressed in equation \ref{eq:normrelation}
  the two curves lie on top of each
  other, and there is no relevant contribution from higher modes.}
\label{fig:sumofmodes}
\end{center}
\end{figure}

Since we only extract QNMs up to $\ell=4$ we need to test whether there is a
relevant contribution from higher modes that we do not extract explicitly. In
the absence of $\ell>4$ modes, summing up all extracted modes up to $\ell=4$
we should recover the full field, according to Eq.~(\ref{eq:normrelation}).
The result of this test for a spinning black hole with $j=0.9$ is shown in
Fig.~\ref{fig:sumofmodes}: at the level of accuracy needed in the present
work, extracting modes with $\ell\leq 4$ is sufficient.

%%%%%%%%%%%%%%%%%%%%%%%%%%%%%%%%%%%%%%%%%%%%%%%%%%%
\subsection{Relative amplitude of corotating and counterrotating modes}
%%%%%%%%%%%%%%%%%%%%%%%%%%%%%%%%%%%%%%%%%%%%%%%%%%%
\label{sec:corot}

We know that when the solution is in the quasinormal ringing regime, it will
behave according to Eq.~(\ref{expansion}). In the previous subsection we have
verified through our simulations the values predicted in Ref.
\cite{Berti:2006wq} for the frequencies. We now also want to verify the
amplitudes of each mode, as predicted in that same reference.

Assume that the observer and the initial data are located far away from the
black hole (these assumptions underlie the ``asymptotic approximation''
adopted in \cite{Andersson95b,Berti:2006wq}).  From Eq.~(4.15) of
\cite{Berti:2006wq}, when $B=0$ the response of the black hole in the ringdown
phase should be well approximated by a QNM decomposition of the form
\be\label{asympt}
\Phi_{\ell m} (r,t)\approx 
-{r_0\over r}\sqrt{\pi} \sigma
\mathcal{R}\left\{
\sum_{n=0}^\infty (\ii A \omega_{\ell mn}) B_{\ell mn} 
e^{-\sigma^2\omega_{\ell mn}^2/4}
e^{-\ii \omega_{\ell mn}(t-r_0-r_*)}
\right\}\,,
\ee
In our simulations we set $A=0$, in which case it can easily be
shown that the previous expression becomes
\be\label{asymptB}
\Phi_{\ell m} (r,t)\approx 
-{r_0\over r}\sqrt{\pi} \sigma B
\mathcal{R}\left\{
\sum_{n=0}^\infty B_{\ell mn} 
e^{-\sigma^2\omega_{\ell mn}^2/4} 
e^{-\ii \omega_{\ell mn}(t-r_0-r_*)}
\right\}\,,
\ee

With respect to \cite{Berti:2006wq} we added an extra factor $r_0/r$. This is
because Eq.~(4.15) in \cite{Berti:2006wq} refers to the Sasaki-Nakamura
function $X_{\ell m}^{(0)}(r,t)$, which is related to the Teukolsky function
$\Phi_{\ell m}(r,t)$ that we are using in our evolutions by the relation
$X_{\ell m}^{(0)}(r,t)=\left(r^2+a^2\right)^{1/2} \Phi_{\ell m}(r,t)$ (see the
discussion in Appendix C of \cite{Berti:2006wq}).  We are interested in large
values of $r$, for which the asymptotic approximation holds and $X_{\ell
  m}^{(0)}(r,t)\simeq r \Phi_{\ell m}(r,t)$.  The transformation between the
Teukolsky and Sasaki-Nakamura functions must also be taken into account when
comparing the initial data in Eq.~(4.14) of \cite{Berti:2006wq} with our
initial data, Eq.~(\ref{eq:initialdata}).  Assuming $\sigma \ll r_0$ and $r\gg
1$ this comparison yields the normalization factor $r_0$ in the equations
above.

The scalar QNM frequencies $\omega_{\ell mn}$ and the scalar excitation
factors $B_{\ell mn}$ are listed in Table I and Table III of
\cite{Berti:2006wq}, respectively. In that reference and in Eq.~(\ref{asympt})
Boyer-Lindquist coordinates are used; since in our simulations we use
Kerr-Schild coordinates we need to transform Eq.~(\ref{asympt}) appropriately.
Since $\Phi$ is a scalar, the transformation is straightforward. The
transformation of the initial data is more subtle, since the slices are
different. One would expect that whenever the asymptotic approximation is
valid the difference between the slices should not be too important. The
results discussed below and explicit comparisons between evolutions using both
coordinate systems in the non-spinning case \cite{Dorband06} confirm this
expectation.  Details on how we transform the initial data and the field
itself are given in Appendix~\ref{sec:coordinatetransform}.

To check the accuracy of Eq.~(\ref{asymptB}), in the rest of this section we
analyze evolutions of different initial data sets, all of them consisting of a
combination of $(\ell=2,m=2)$ and $(\ell=2,m=-2)$ modes with $A=0$ and $B=1$.
We numerically explore the dependence of the amplitudes of the counter- and
co-rotating {\em fundamental} modes (in the next subsection we will study
overtones) on the width $\sigma$ of the initial data [cf.\
Eq.~(\ref{eq:initialdata})]. In order to assess more quantitatively the effect
of the {\it time-shift problem} (see Sec.~\ref{timeshiftproblem}) we first
compare the value of the width maximizing these amplitudes.  Given that all
the initial data sets that we consider are centered at the same radius, we can
make the reasonable assumption that locally (that is, around the width for
which the amplitudes are maximal) $t_0$ is approximately the same for each
set. If $t_0$ were {\em exactly} the same, the value of $t_0$ used would not
change the width at which the maximum amplitude is located, since changes in
$t_0$ would only involve a global rescaling of all amplitudes, as discussed in
Sec.~\ref{timeshiftproblem}.  Therefore the hope is that within the setting
described for our simulations the width for which the amplitudes are maximal
does not depend too sensitively on $t_0$.

The numerical results shown here were obtained with the same number of points
in the angular direction as above. We used half the resolution in the radial
direction (that is, $\Delta r = M/5$) for a rough scan of a large $\sigma$
range, and again the original resolution around the maxima of the amplitudes.
We chose initial data with varying widths $\sigma$, $r_0=20M$
(as in the simulations above) and an observer at $r=40M$, for which the
asymptotic approximation holds reasonably well \cite{Dorband06}.  We picked
$t_{0}=r_{\mathrm{initial\ data}}+r$ (that is, $t_0=60M$ in the
cases considered), which is approximately the time the initial data pulse
needs to propagate to the black hole and back to the observer.

\begin{figure}[htbp]
\begin{center}
\includegraphics{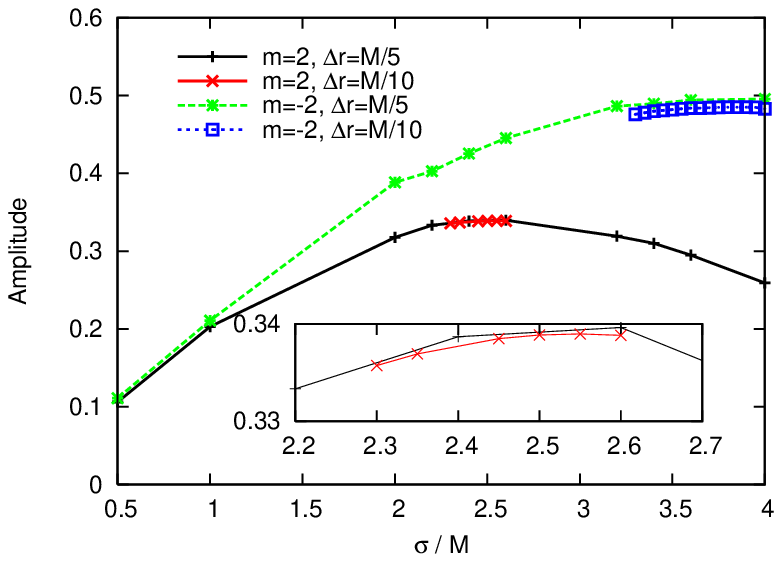}
\includegraphics{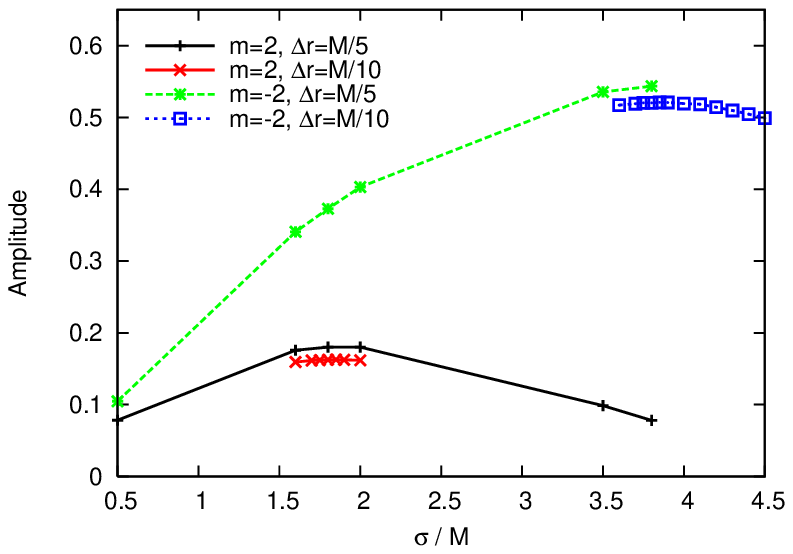}
\caption{Numerically obtained excitation amplitudes of the $\ell=|m|=2$
  fundamental modes assuming an observer location $r_{\mathrm{obs}}=40M$ and a
  ringdown starting time $t_{0}=60M$. The left panel refers to a black hole
  with spin $j=0.5$. According to predictions from perturbation theory in the
  asymptotic approximation [cf.  Eq.~(\ref{asympt}) and the following
  discussion] the maximum for $m=2$ should be located at $\sigma_{220}=2.445$,
  while the value that we obtain from our simulations is $\sigma_{220}=2.55\pm
  0.05$ (the uncertainty describing the difference between consecutive values
  of $\sigma$ used in our simulations: $\Delta \sigma=0.05$).  Similarly, for
  $m=-2$ the width at the maximum should be $\sigma_{2-20}=3.434$, while we
  obtain $\sigma_{2-20}=3.875\pm 0.075$.  The right panel, in turn, refers to
  a black hole with spin $j=0.9$. In this case the theoretical (numerical)
  maxima are located at $\sigma_{220}=1.816$ ($\sigma_{220}=1.85\pm 0.05$) and
  $\sigma_{2-20}=3.758$ ($\sigma_{2-20}=3.85\pm 0.05$), respectively. The
  inset in the left panel is a zoom around the maximum for $j=0.5$ and $m=2$.
  As discussed in the text, an uncertainty in the excitation time of $0.09M$
  would already explain the difference between the predicted location of the
  maxima and our numerical results.}
\label{fig:amplitudes}
\end{center}
\end{figure}

\begin{table}[htdp]
  \caption{Excitation amplitudes for $j=0.5$, $\ell=2$ and $n=0$ for initial
    perturbations of variable Gaussian width $\sigma$, as displayed in
    Fig.~\ref{fig:amplitudes}. The observer location in these runs is
    $r=40M$. Highlighted are the maxima in the amplitudes of
    the different $m$-modes.  Also shown are the relative amplitudes of the
    two modes, and the relative differences between the values predicted by
    perturbation theory and the ones extracted from our numerical
    simulations. The amplitudes are given for the wave 
    expressed in Boyer Lindquist coordinates
    (see appendix \ref{sec:coordinatetransform} for details) and are multiplied by a factor of
    $r/r_0$ to get them in an observer independent form.}
\begin{center}
\begin{tabular}{| c | c | c | c || c | c | c | c |}
\hline
 & \multicolumn{3}{|l||}{numerical results} &
 \multicolumn{3}{|l|}{perturbation theory} & relative difference\\
 \hline
$\sigma$ & ${\cal A}_{220}$ & ${\cal A}_{2-20}$ & ${\cal A}_{2-20}/{\cal A}_{220}$ & 
${\cal A}_{220}$ & ${\cal A}_{2-20}$ & ${\cal A}_{2-20}/{\cal A}_{220}$ & ${\cal A}_{2-20}/{\cal A}_{220}$ \\
\hline \hline
      2.30 &     0.3357 &      0.411 &       1.22 &      0.315 &      0.419 &       1.33 &      0.080 \\

      2.35 &     0.3369 &      0.410 &       1.22 &      0.314 &      0.423 &       1.35 &      0.097 \\

      2.45 &     0.3385 &      0.420 &       1.24 &      0.311 &      0.430 &       1.38 &      0.103 \\

      2.50 &     0.3389 &      0.425 &       1.25 &      0.310 &      0.433 &       1.40 &      0.102 \\

      {\bf 2.55} &     {\bf 0.3390} &      0.430 &       1.27 &      0.308 &      0.436 &       1.42 &      0.104 \\

      2.60 &     0.3388 &      0.435 &       1.28 &      0.306 &      0.439 &       1.43 &      0.105 \\

      3.30 &      0.315 &      0.476 &       1.51 &      0.253 &      0.449 &       1.77 &      0.149 \\

      3.35 &      0.311 &      0.478 &       1.54 &      0.248 &      0.448 &       1.81 &      0.149 \\

      3.40 &      0.308 &      0.480 &       1.56 &      0.243 &      0.446 &       1.84 &      0.151 \\

      3.45 &      0.304 &      0.481 &       1.58 &      0.238 &      0.445 &       1.87 &      0.154 \\

      3.50 &      0.301 &      0.482 &       1.60 &      0.233 &      0.443 &       1.90 &      0.158 \\

      3.55 &      0.297 &      0.483 &       1.63 &      0.228 &      0.441 &       1.93 &      0.159 \\

      3.60 &      0.293 &      0.484 &       1.65 &      0.223 &      0.439 &       1.97 &      0.161 \\

      3.65 &      0.289 &      0.484 &       1.67 &      0.217 &      0.437 &       2.01 &      0.168 \\

      3.70 &      0.285 &     0.4843 &       1.70 &      0.212 &      0.434 &       2.05 &      0.170 \\

      3.75 &      0.281 &     0.4848 &       1.73 &      0.207 &      0.432 &       2.09 &      0.173 \\

      3.80 &      0.276 &     0.4850 &       1.76 &      0.202 &      0.429 &       2.12 &      0.173 \\

      {\bf 3.85} &      0.272 &     {\bf 0.4851} &       1.78 &      0.196 &      0.425 &       2.17 &      0.178 \\

      {\bf 3.90} &      0.267 &     {\bf 0.4851} &       1.82 &      0.191 &      0.423 &       2.21 &      0.180 \\

      3.95 &      0.263 &     0.4849 &       1.84 &      0.186 &      0.419 &       2.25 &      0.182 \\

      4.00 &      0.259 &     0.4831 &       1.87 &      0.180 &      0.416 &       2.31 &      0.193 \\\hline
\end{tabular}
\label{tab:amp1}
\end{center}
\end{table}

\begin{table}[htdp]
\caption{Same as Table \ref{tab:amp1} for $j=0.9$.}
\begin{center}
\begin{tabular}{| c | c | c | c || c | c | c | c |}
\hline
 & \multicolumn{3}{|l||}{numerical results} &
 \multicolumn{3}{|l|}{perturbation theory} & relative difference\\
 \hline
$\sigma$ & ${\cal A}_{220}$ & ${\cal A}_{2-20}$ & ${\cal A}_{2-20}/{\cal A}_{220}$ & 
${\cal A}_{220}$ & ${\cal A}_{2-20}$ & ${\cal A}_{2-20}/{\cal A}_{220}$ & ${\cal A}_{2-20}/{\cal A}_{220}$ \\
\hline \hline

      1.60 &     0.1594 &     0.3156 &       1.98 &     0.1768 &     0.3683 &       2.08 &       0.05 \\
      1.70 &     0.1615 &     0.3319 &       2.06 &     0.1766 &     0.3857 &       2.18 &       0.06 \\
   1.75 &     0.1621 &     0.3399 &       2.10 &     0.1755 &     0.3990 &       2.27 &       0.08 \\
     1.80 &     0.1625 &     0.3476 &       2.14 &     0.1752 &     0.4022 &       2.30 &       0.07 \\
      {\bf 1.85} &  {\bf 0.1626} &     0.3553 &       2.18 &     0.1740 &     0.4101 &       2.36 &       0.07 \\
     1.90 &     0.1625 &     0.3629 &       2.23 &     0.1725 &     0.4177 &       2.42 &       0.08 \\
     2.00 &     0.1617 &     0.3775 &       2.33 &     0.1725 &     0.4323 &       2.51 &       0.07 \\
      3.60 &     0.0800 &     0.5173 &       6.47 &     0.0566 &     0.5259 &       9.30 &       0.30 \\
      3.70 &     0.0743 &     0.5192 &       6.99 &     0.0507 &     0.5235 &      10.32 &       0.32 \\
     3.75 &     0.0714 &     0.5204 &       7.29 &     0.0479 &     0.5220 &      10.89 &       0.33 \\
    3.80 &     0.0688 &     0.5204 &       7.56 &     0.0452 &     0.5203 &      11.51 &       0.34 \\
     {\bf 3.85} &     0.0661 &     {\bf 0.5212} &       7.89 &     0.0427 &     0.5184 &      12.15 &       0.35 \\
      3.90 &     0.0636 &     0.5208 &       8.19 &     0.0402 &     0.5162 &      12.85 &       0.36 \\
     4.00 &     0.0590 &     0.5194 &       8.81 &     0.0355 &     0.5114 &      14.41 &       0.39 \\
    4.10 &     0.0590 &     0.5184 &       8.79 &     0.0312 &     0.5059 &      16.20 &       0.46 \\
     4.20 &     0.0505 &     0.5144 &      10.19 &     0.0274 &     0.4997 &      18.25 &       0.44 \\
   4.30 &     0.0469 &     0.5098 &      10.88 &     0.0239 &     0.4929 &      20.64 &       0.47 \\
     4.40 &     0.0433 &     0.5047 &      11.65 &     0.0207 &     0.4854 &      23.41 &       0.50 \\
    4.50 &     0.0433 &     0.4991 &      11.52 &     0.0179 &     0.4774 &      26.63 &       0.57 \\
\hline
\end{tabular}
\label{tab:amp2}
\end{center}
\end{table}

Figure \ref{fig:amplitudes} and (more quantitatively) Tables \ref{tab:amp1}
and \ref{tab:amp2} show the excitation amplitudes as functions of the width of
the Gaussian $\sigma$ from our numerical simulations. At first our results
could be interpreted as an approximate verification of the predictions of
\cite{Berti:2006wq}. However, if one takes into account the limitations
imposed by the {\it time-shift problem}, the agreement can in fact be
considered excellent.  For example, take the $j=0.5, m=-2$ case, which is the
one where the difference between the theoretical and numerical values is
largest. The theoretical maximum is located at $\sigma = 3.434M$, while the
numerical value is $\sigma = (3.875\pm 0.075)M$ (the uncertainty indicating
the difference between consecutive values of $\sigma$). The relative numerical
amplitude between $\sigma=3.45M$ and $\sigma=3.85M-3.9M$ from our simulations
is $\approx 1.008$ (see Table \ref{tab:amp1}). If the values of $t_0$ for
these two widths differ by $\approx 0.09M$, the amplitude corresponding to
$\sigma=3.45M$ would actually be larger than the one of $\sigma=3.85M-3.9M$
and would therefore shift the maximum to the predicted value of $3.45M$.
Recalling that we used $t_0=60M$, a very modest uncertainty in the {\em
  relative} ringdown starting time ($\approx 0.4\%$) would shift the maximum
to the theoretical value.  We also assumed the same excitation time $t_0$ for
all the initial data sets when fitting our numerical data. Whenever such
assumption is a good approximation, the precise value of $t_0$ should not
affect the location of the width for which the excitation amplitudes is
maximal. In particular, the approximation should be good if the initial data
pulses are relatively narrow.  However, as $\sigma$ increases, the possibility
of the excitation time $t_0$ shifting around has to be taken into account,
because the interaction time of the pulse with the black hole becomes longer
and the interaction sets in well before the center of the pulse reaches the
black hole.  Taking all this into account, the agreement between numerical and
perturbative results {\em for the location of the maxima} can be considered
excellent. The situation for the amplitudes themselves is different, as
discussed next.

Tables \ref{tab:amp1} and \ref{tab:amp2} show the predicted and extracted
absolute and relative amplitudes for the co- and counter-rotating modes ${\cal
  A}_{m=2}$, ${\cal A}_{m=-2}$.  As expected, the prediction from perturbation
theory works better for sharp pulses. The differences between the predicted
and absolute values are of order a few percent for sharp pulses and grow with
$\sigma$.  For $\sigma=4$ the difference is as large as $\sim 20 \%$ and $\sim
60\%$ for $j=0.5$ and $j=0.9$, respectively (the actual amplitudes being
larger than the predicted ones).  These large differences in the relative
amplitudes are mostly due to the amplitude of the corotating mode, the
predicted and extracted amplitudes for the counterrotating one agree quite
well. The fact that the location of the maxima, as discussed above, agrees
very well despite the large differences in the amplitudes for larger $\sigma$
can be easily explained: the location of the maxima for the corotating mode
takes place at $\sigma \approx 1.85M$, which corresponds to a pulse which is
sharp enough for perturbation theory to give a good prediction, while the
maximum for the counterrotating mode is at a larger value of $\sigma$ but, as
we have discussed, the agreement between predicted and measured amplitudes is
quite good for that mode.

Could this large difference in amplitudes be explained by the {\it time-shift problem}, 
as discussed in Sec.~\ref{timeshiftproblem}? Using
Eq.~(\ref{eq:timeshiftrelamp}) and assuming that $t_0$ is roughly the same for
both modes we find that an uncertainty in the excitation time as large as
$\delta_0=\pm 5M$ would imply an uncertainty on the relative amplitudes of
about $\pm 1.1\%$ for $j=0.5$, and $\pm 13\%$ for $j=0.9$. Therefore the
uncertainty $\delta_0$ does not seem to account for the differences that we
find with respect to the predicted amplitudes.  One possibility is that the
excitation time $t_0$ is different for the two modes in a pair; but, if so, it
is not clear then why our naive choice of $t_0$ is very good for the
counterrotating mode and quite bad for the corotating one.  It is actually not
clear why such a large disagreement happens only for the corotating mode, and
not for the counterrotating one.  The possibilities that the initial data
and/or the observer are not far enough away for the asymptotic approximation
to be valid, or that the disagreement is due to a lack of resolution, seem to
be ruled out by one-dimensional studies in the non-spinning case
\cite{Dorband06}.  Summarizing, even though the exact mechanism is not clear,
all this suggests that the predicted amplitudes for the corotating mode in the
asymptotic approximation are simply valid only for very sharp pulses, as the
black hole spin increases.

To conclude, we want to discuss one aspect of our simulations, as shown in
Figure \ref{fig:amplitudes}. We see a rather large discrepancy between the
amplitudes resulting from runs with resolution $\Delta r=M/5$ and $\Delta r =
M/10$, especially for $j=0.9$. That is a direct effect of decreasing accuracy
in $\mathcal{I} (\omega_{\ell mn})$ when going to high spins (see
Sec.~\ref{sec:fittingprocedure}) and the need for more resolution in those
cases. The {\em location} of the maximum, however, is always consistent (that
is, within the differences in $\sigma$ used in the different initial data
sets) between runs of different resolution. That is not surprising since the
measured $\omega_{\ell mn}$ at a fixed resolution is roughly the same for all
values of $\sigma$, and the value of $\sigma$ that maximizes ${\cal A}_{\ell
  mn}$ only depends on the value of $\omega_{\ell mn}$.

%%%%%%%%%%%%%%%%%%%%%%%%%%%%%%%%%%%%%%%%%%%%%%
\subsection{Overtones and rapidly spinning black holes}
\label{overtones}
%%%%%%%%%%%%%%%%%%%%%%%%%%%%%%%%%%%%%%%%%%%%%%

As discussed in the introduction, a single complex quasinormal frequency
contains enough information to determine the two parameters of a Kerr black
hole (namely, its mass $M$ and spin $j$). If one is able to detect a second
mode from the same source, one can use this extra information for a
consistency check that would increase the confidence in the interpretation of
the measured data as signals from a perturbed black hole.  An important
question that might be answered by numerical relativity is whether more than
one mode will be detectable by Earth- and space-based gravitational wave
detectors. In Sec.~\ref{sec:corot} we considered the relative amplitude of
corotating and counterrotating modes; here we use our simulations to determine
the relative excitation of overtones with the same angular dependence and
$m>0$.  According to perturbation theory, in this case the damping time of the
first overtone becomes comparable to the damping time of the fundamental mode
for large spins (see Fig.~\ref{fig:delta0}). In addition, the excitation
factor of higher overtones is usually larger than the excitation factor of the
fundamental mode for large $j$ \cite{Berti:2006wq}. This means that higher
overtones are more likely to be detectable for fast spinning black holes. A
detailed study of this topic is beyond the scope of this paper, but here we
briefly discuss how we can extract information about overtones from our data
and determine which modes contribute most significantly to the waveform.

We perform simulations for different spins ($j=0$, $0.5$, $0.9$, and $0.98$).
The initial data and numerical procedure are the same as in Sec.~\ref{initial}
and \ref{specs}, with one exception: for spins $j\geq 0.9$ we found it
necessary to increase the angular resolution. The simulations presented in
this section used a resolution of $31\times31$ grid points on each block in
the angular directions. This is not surprising, since for fast rotation we
expect more dynamics in the angular directions.

The extraction of modes is done in principle according to
Sec.~\ref{sec:fittingprocedure}.  Extracting information about all modes
present in the data can turn into a subtle problem, especially when the
contributions of some modes is weak. One option is to first fit for the
strongest mode present in the data, subtract the fit, fit for the next
dominant mode and so on, repeating the procedure as long as an oscillatory
exponential decay is seen in the data. However, when there are several modes
with similar contributions we can just fit for all of them at the same time.
This is exactly what we did for fundamental modes with different $m$ in the
previous subsection. When a single mode dominates the waveform the first
strategy not only seems to be more meaningful, but also turns out to work
better in practice. The results of this section were computed by a hybrid of
these two methods, depending on the contribution of each mode (something that
one can find out by, for example, looking at the dominant frequencies of the
signal to fit).

Table \ref{tab:overtones} shows the quasinormal frequencies of the overtones
that we get from our simulations, using $(A=0,B=1)$, $\sigma=M, r_0=20M$ and 
an observer at
$r=60M$. We find that the overtones for the $m=-2$ mode do not
contribute enough to the waveforms to extract them from our data with decent
accuracy, especially for high spins.  The reason for this is that the
imaginary part of their frequency is generally smaller than the one for the
corresponding $m=2$ mode, which makes them decay faster.  The decay of the
$m=2$ mode, on the other hand, slows down considerably when increasing the
spin. We numerically find that the excitation amplitude (at fixed $t_0$)
increases with increasing spin. Those two effects combined make the extraction
of overtones easier and more accurate in the high spin cases. Quite
remarkably, for runs with spin $j=0.9$ and above we can extract the
quasinormal frequency for $n=2$ with reasonable accuracy (see Table
\ref{tab:overtones}).

Table \ref{amps_overtones} compares the amplitudes of the three most dominant
$l=2$ modes, ($m=2, n=0$), ($m=-2, n=0$) and ($m=2, n=1$), with the predicted
asymptotic amplitudes of Eq.~(\ref{asympt}). Except for the $j=0.98$ case, the
difference between the predicted and extracted values for the relative
amplitudes between a given mode and the fundamental $\ell=2=m$ one is of the
order of a few percent for the fundamental mode and one order of magnitude
larger for the first overtone. 
 
\begin{table}[htdp]
\begin{center}
  \caption{Comparison of quasinormal frequencies for the first overtones
    ($n=1,2$) of an $\ell=2$, $m=2$ mode, for black holes with varying spin,
    as predicted by perturbation theory and as extracted from our numerical
    simulations, along with their relative differences. This table is
    complementary to Table \ref{tab:freq}, where we show the frequencies
    associated to the fundamental modes. The extraction of overtones becomes
    easier for rapidly rotating black holes, as explained in the text,
    allowing us to extract the frequencies of two overtones for high spins. }
\label{tab:overtones}
\begin{tabular}{| l | l | l | l | l | l |}
\hline
$j$ & $n$ & $\omega_{\mathrm{perturb}}$ & $\omega_{\mathrm{numerical}}$ & rel.\ difference (Re, Im) \\
\hline \hline
0.0 & $1$ & $0.46385-0.29560\ii$ & $0.45651-0.28859\ii$ & $1.58\times10^{-2}$, $2.37\times10^{-2}$  \\
\hline
0.5 & $1$ & $0.57344-0.28334\ii$ & $0.54718-0.31722\ii$ & $4.58\times10^{-2}$, $1.20\times10^{-1}$ \\
\hline
0.9 & $1$  & $0.77768- 0.20801 \ii$  & $0.73737- 0.19558 \ii$ &  $5.18\times10^{-2}$, $5.98\times10^{-2}$\\
0.9 & $2$ & $0.77043- 0.34720 \ii$  & $0.52473- 0.35319 \ii$ &  $3.19\times10^{-1}$,  $1.73\times10^{-2}$\\
\hline
0.98 & $1$ & $0.89622- 0.12214 \ii$  & $0.93152- 0.12406 \ii$ &  $3.94\times10^{-2}$,  $1.57\times10^{-2}$\\
0.98 & $2$  & $0.89358- 0.20244 \ii$  & $0.88668- 0.25850 \ii$ &  $7.72\times10^{-3}$, $2.77\times10^{-1}$\\
\hline
\end{tabular}
\end{center}
\end{table}%

\begin{table}
\begin{center}
  \caption{Absolute and relative amplitudes as a function of the black hole
    spin and angular dependence of the perturbations, as predicted by
    perturbative calculations and as extracted from our numerical evolutions.
    The amplitudes are given for the wave 
    expressed in Boyer Lindquist coordinates
    (see appendix \ref{sec:coordinatetransform} for details) and are multiplied by a factor of
    $r/r_0$ to get them in an observer independent form.
    The last column presents the relative difference between perturbative and
    numerical results for relative amplitudes. In the corotating case we also
    extract the amplitude of the first overtone. The differences in the
    relative amplitudes are considerably smaller when we look at corotating
    and counterrotating modes, compared to the case of fundamental mode and
    first overtone with the same angular dependence. This can be explained by
    the relative magnitude of their damping frequencies, as discussed in
    Section \ref{timeshiftproblem} (see also Table \ref{tab:overtones}). This
    difference becomes less pronounced at very large spins, as expected from
    the analysis of Section \ref{timeshiftproblem}. \label{amps_overtones}}
\begin{tabular}{| l | r r r || l | l || l | l | l |}
\hline
&\multicolumn{3}{|l||}{mode} & \multicolumn{2}{l ||}{numerical result}
& \multicolumn{2}{l |}{perturbation theory} & relative difference\\
\hline
  $j$ & $l$ & $m$ & $n$ &    ${\cal A}_{\ell mn}$ &  ${\cal A}_{\ell mn}/{\cal
    A}_{220}$ &  ${\cal A}_{\ell mn}$ &  ${\cal A}_{\ell mn}/{\cal
    A}_{220}$ &  
 ${\cal A}_{\ell mn}/{\cal A}_{220}$\\
  \hline \hline
0.00 &  2 & 2 & 0 &      0.211 &       1.00 &      0.221  &       1.00
& 0.00\\ 
0.00 &  2 & 2 & 1 &      0.316 &       1.50 &      0.504 &       2.28
& 0.342 \\
\hline 
0.50 &  2 & 2 & 0 &      0.201 &       1.00 &      0.213 &       1.00
 & 0.00 \\
0.50 & 2 & -2 & 0 &      0.208 &       1.03 &      0.228 &       1.07
& 0.037\\
0.50 & 2 & 2  & 1 & 	   0.525 &		 2.61 &		0.768 &		   3.61
& 0.277 \\
\hline
0.90 &  2 & 2 & 0 &      0.137 &       1.00 &      0.148 &       1.00
& 0.00 \\
0.90 & 2 & -2 & 0 &      0.211 &       1.54 &      0.246 &       1.66
& 0.072 \\ 
0.90 &  2 & 2 & 1 &      0.533 &       3.89 &       0.98 &       6.62
& 0.412\\
\hline
0.98 &  2 & 2 & 0 &     0.0833 &       1.00 &      0.068 &       1.00
&  0.00 \\
0.98 & 2 & -2 & 0 &      0.263 &       3.16 &      0.257 &       3.78
 & 0.164\\
0.98 &  2 & 2 & 1 &      0.634 &       7.61 &      0.416 &       6.12
& 0.243 \\
\hline
\end{tabular}  
\end{center}
\end{table}

%%%%%%%%%%%%%%%%%%%%%%%%%%%%%%%%%%%%%%%%%%%%%
\section{Conclusions}
%%%%%%%%%%%%%%%%%%%%%%%%%%%%%%%%%%%%%%%%%%%%%

The chances of a multi-mode detection by either Earth- or space-based
gravitational wave detectors will depend on the relative amplitude of those
modes. Knowing in advance which modes should be excited under a realistic
binary merger would reduce the dimensionality of the template bank needed to
perform matched filtering on ringdown waveforms. An answer that numerical
relativity might provide is precisely which modes are likely to be dominant.
This involves predicting the relative amplitudes of different pairs of modes
under a variety of scenarios. In this paper we have taken a first step towards
understanding the issues involved in such a prediction.

We first presented a systematic way of extracting QNMs from a given signal.
Our procedure has a number of built in self-consistency checks, to make sure
that when we keep adding modes to our fit we are fitting a true signal and not
numerical noise. One of these self-consistency checks is to make sure that we
extract the correct quasinormal frequency of each mode within a certain
accuracy. If the data being analyzed comes from a numerical simulation,
consistent frequencies can be used to monitor the accuracy of the code. If the
data is experimental, consistency of the frequencies allows for a test of the
no hair conjecture. In more detail: during our fitting procedure we first fit
for the dominant mode(s), look at the residual (defined as the difference
between our original signal and the fit), make sure that it has a consistent
quasinormal ringing behavior and only then fit for the next set of modes,
repeating the procedure as long as it makes sense to do so. By following this
procedure we have hardly been able to go beyond the first few dominant modes,
and this was only possible in very special cases. We expect this to happen
with most numerical simulations.

We addressed in some detail the so-called {\it time-shift problem}. In essence,
this is the fact that the quasinormal amplitudes depend exponentially on the
quasinormal ringing excitation time, which is not defined unambiguously (not
even in the continuum). Furthermore, examining actual values of quasinormal
frequencies we have seen that this exponential dependence is an important
factor to take into account in practice.  To (partially) get rid of this
exponential dependence we propose to look at relative amplitudes: choosing
pairs of modes whose damping frequency is as close as possible, we can
partially cancel each others exponential dependence. We analyzed in detail
the exponential dependence of different pairs of modes as a function of the
black hole spin.  In particular, we found that the {\it time-shift problem} becomes
more important as one increases the spin. For modes with the same value of
$\ell$, for example, the problem is not very relevant for spins $j \lesssim
0.5$. On the other hand, an accurate extraction of the relative amplitude
between the fundamental mode and the first overtone only seems feasible for
very high spins and $m>0$.

Keeping this in mind, we first extracted the fundamental quasinormal
frequencies for different values of spin, ranging from $j=0$ to a rapidly
rotating black hole with $j=0.98$. Even using modest resolutions our
frequencies agree with those obtained from perturbation theory within one part
in $10^5$ to one part in $10^2$, depending on the black hole spin, location of
the observer and angular dependence. To our knowledge 
this is the first time that quasinormal frequencies for
scalar perturbations of Kerr, as predicted by perturbation theory, have been
verified by numerical evolutions of the field equations.

Next we analyzed in detail the relative amplitude of corotating and
counterrotating fundamental modes, as a function of the width of the initial
perturbation and the black hole spin, being able to quantify (within the limitations
imposed by the {\it time-shift problem}) under what conditions the
asymptotic approximation of Ref.~\cite{Berti:2006wq} is valid. In particular, we were able to verify the widths
of the initial perturbation corresponding to the maximal QNM excitation.
Finally, we studied the excitation of overtones. We found that, according to
expectations from perturbation theory \cite{Berti:2006wq}, they get
significantly excited for corotating modes and very high spins. In this
particular case we were able to extract the complex QNM frequency for the
fundamental mode and the first two overtones, with a difference with respect
to the predicted values by perturbation theory of the order of a tenth of a
percent to ten percent, depending on the mode and the black hole spin.
We expect the techniques and results of this paper to be general enough to
be useful for future work on ringdown waveforms.

%%%%%%%%%%%%%%%%%%%%%%%%%%%%%%%%%%%%%%%%%%%%%
\begin{acknowledgments}
%%%%%%%%%%%%%%%%%%%%%%%%%%%%%%%%%%%%%%%%%%%%%

  This research was supported in part by NSF grant PHY 05-05761, NASA grant
  NAG5-1430, and the National Center for Supercomputer Applications grant
  MCA02N014 to Louisiana State University, and by NSF grant PHY 03-53180 and
  NASA Grant NNG06GI60 to Washington University. It also employed the
  resources of the Center for Computation and Technology at Louisiana State
  University, which is supported by funding from the Louisiana legislature's
  Information Technology Initiative. We are particularly grateful to Vitor
  Cardoso for many helpful suggestions and discussions. M. Tiglio thanks S.
  Teukolsky for hospitality at Cornell University, where part of this work was
  done. We thank L. Lehner for helpful comments on the manuscript. 
  Our numerical calculations used the Cactus framework
  \cite{Goodale02a,cactusweb1} with a number of locally developed thorns, the
  LAPACK \cite{lapackweb} and BLAS \cite{blasweb} libraries from the Netlib
  Repository \cite{netlibweb}, and the LAM \cite{burns94:_lam,
    squyres03:_compon_archit_lam_mpi, lamweb} and MPICH \cite{Gropp:1996:HPI,
    mpich-user, mpichweb} MPI \cite{mpiweb} implementations.
\end{acknowledgments}

%%%%%%%%%%%%%%%%%%%%%%%%%%%%%%%%%%%%%%%%%%%%%%%
\appendix
%%%%%%%%%%%%%%%%%%%%%%%%%%%%%%%%%%%%%%%%%%%%%%%
%%%%%%%%%%%%%%%%%%%%%%%%%%%%%%%%%%%%%%%%%%%%%
\section{Change of coordinates and initial data}
\label{sec:coordinatetransform}
%%%%%%%%%%%%%%%%%%%%%%%%%%%%%%%%%%%%%%%%%%%%%
To compare the numerical results with predictions from perturbation theory we
must switch from the usual Boyer-Linquist coordinates (as used, for example,
in \cite{Berti:2006wq}) to the Kerr-Schild coordinates used in our code.  We
denote by ($r_*$, $t$) the Boyer-Lindquist radial tortoise coordinate and
time, and by ($\bar r$, $\bar t$) the Kerr-Schild coordinates.  The transformation we
need is given by
\begin{eqnarray}
\label{eq;coordinatetransformation}
% Without spin
% t(\bar t,\bar r) &=& \bar t - 2 \ln \left( {\bar r\over2}-1 \right) + \tilde t\\
% r_*(\bar r) &=& \bar r + 2 \ln \left( {\bar r\over2}-1 \right).
% With spin
t(\bar t,\bar r) &=& \bar t  - \Omega(\bar r) + \tilde t\,,\\
r_*(\bar r) &=& \bar r + \Omega(\bar r)\,,
\end{eqnarray} 
with the definitions
\begin{eqnarray}
\Omega(r) & = & {2 M r_+ \over r_+-r_-} \ln \left( {r - r_+ \over 2 M} \right) - 
	{2 M r_- \over r_+-r_-} \ln \left( {r - r_- \over 2 M} \right)\,, \\
r_+ &=& M + \sqrt{M^2-a^2}\,, \\
r_- &=& M - \sqrt{M^2-a^2}\,. \\
\end{eqnarray}
The reference time $\tilde t$ can in principle be freely chosen and is used to
define where $t(\bar r,\bar t)$ crosses zero. We fix it by the condition that in both
coordinate systems the initial pulse is at the same physical
distance from the black hole, i.e. $t(\bar t=0, \bar r=\bar r_0) = 0$:
\begin{equation}
	\tilde t = \Omega(\bar r_0)\,.
\end{equation}
The location of the initial pulse in these coordinates becomes
\begin{eqnarray}
r_0 &=& \bar r_0 + \Omega(\bar r_0).
\end{eqnarray} 
For consistency, the value of $\sigma$ has to be adjusted to 
 tortoise coordinates. As a rough approximation we set
\begin{equation}
\label{eq:sigmatransformation}
\sigma = {1\over2} \left[r_*(\bar r+\bar \sigma) - r_*(\bar r-\bar \sigma)\right].
\end{equation}

From equations \ref{expansion} and \ref{asymptB} we read of the
exponential decay of each modes amplitude in 
Boyer-Lindquist coordinates and then substitute them by the 
Kerr-Schild coordinates. We choose $t_0=r_*+r_0$.
\begin{equation}
	\mathcal{A}_{\ell mn} e^{t-r_*-r_0} = \mathcal{A}_{\ell mn}  e^{ - 2 \Omega(\bar r) {\cal I} (\omega_{lmn}) } e^{\bar t-\bar r-\bar r_0} \equiv  \mathcal{\bar A}_{\ell mn} e^{\bar t-\bar r-\bar r_0}
\end{equation}
This equation relates the amplitudes as seen in Boyer-Lindquist coordinates $\mathcal{A}_{\ell mn}$
with the ones found in the simulations that were done in Kerr-Schild coordinates $\mathcal{\bar A}_{\ell mn}$.

% Use this for gr-qc
\bibliographystyle{bibtex/apsrev-titles}
% Use this for PRD
%\bibliographystyle{bibtex/apsrev-nourl}

\bibliography{bibtex/references}

\end{document}